\RequirePackage{fix-cm}

\documentclass[twocolumn,epjc3,amssymb,eqsecnum,amsfonts,epsf,preprintnumbers,nofootinbib,superscriptaddress]{svjour3}
\smartqed
\RequirePackage{graphicx}
\usepackage{multirow}
\usepackage{amssymb,graphics}
\usepackage{graphics,epsfig,subfigure}
\usepackage{epstopdf}
\usepackage{graphicx}
\usepackage{dcolumn,color}
\usepackage{bm}
\usepackage{epsfig}
\usepackage{hyperref}
\usepackage{color,framed}
\usepackage{lipsum}
\usepackage{amsmath,amssymb,graphics,epsfig,subfigure,amsfonts,latexsym,color,makeidx}
\usepackage{cite}
\newcommand {\nn}    {\nonumber}

\journalname{Eur. Phys. J. C}

\begin{document}
\renewcommand{\baselinestretch}{1.3}
\title{Evolution of scalar field resonances in a braneworld}

%\author{Qin Tan$^{a}$$^{b}$\footnote{Yu-Peng Zhang and Qin Tan are co-first authors of the article.}}
%\author{Yu-Peng Zhang$^{a}$$^{b*}$}
%\author{Wen-Di Guo$^{a}$$^{b}$}
%\author{Jing Chen$^{a}$$^{b}$}
%\author{Chun-Chun Zhu$^{a}$$^{b}$}
%\author{Yu-Xiao Liu$^{a}$$^{b}$\footnote{liuyx@lzu.edu.cn, corresponding author}}
\author{Qin Tan\thanksref{e1,addr1,addr2}
\and  Yu-Peng Zhang\thanksref{e1,addr1,addr2}
\and  Wen-Di Guo\thanksref{addr1,addr2}
 \and Jing Chen\thanksref{addr1,addr2}
 \and Chun-Chun Zhu\thanksref{addr1,addr2}
\and Yu-Xiao Liu \thanksref{e2,addr1,addr2}}

\thankstext{e1}{Qin Tan and Yu-Peng Zhang are co-first authors of the article.}
\thankstext{e2}{e-mail:liuyx@lzu.edu.cn, corresponding author}

\institute{Lanzhou Center for Theoretical Physics, Key Laboratory of Theoretical Physics of Gansu Province, School of Physical Science and Technology, Lanzhou University, Lanzhou 730000, China\label{addr1}
\and
Institute of Theoretical Physics and Research Center of Gravitation, Lanzhou University, Lanzhou 730000, China \label{addr2}}

\date{}

\maketitle

\begin{abstract}
In this work, we investigate numerical evolution of massive Kaluza-Klein~(KK) modes of a scalar field in a thick brane. We derive the Klein-Gordon equation in five dimensional spacetime, and obtain the evolution equation and the Schr\"odinger-like equation. With the resonances of the scalar KK modes as the initial data, the scalar field is evolved with the  maximally dissipative boundary condition. The results show that there are scalar KK resonant particles with long life on the brane, which indicates that these resonances might be viewed as one of the candidates for dark matter.
\end{abstract}

\section{Introduction}
\label{Introduction}
The nature of dark matter (DM) constitutes one of the most long-standing and puzzling questions in cosmology. There is abundant evidence that nonluminous matter makes up a large fraction of all matter in our universe. Results from cosmological measurements have now determined with exquisite precision the abundance of DM~\cite{Planck:2018vyg}. But the identity of DM remains a mystery. Recently, Barranco $et~al.$ proposed that the scalar dynamical resonances can form long-lived configurations around black holes~\cite{Barranco:2011eyw,Barranco:2012qs,Barranco:2013rua}. This ultra-light scalar resonances might be regarded as one of the candidates for DM. Then, such long-lived distribution was generalized to the Dirac field by Zhou $et~al.$~\cite{Zhou:2013dra}. After that, these massive dynamical resonances around black holes attracted a lot of interest \cite{Gossel:2013cka,Decanini:2014bwa,Sampaio:2014swa,Degollado:2014vsa,
Sanchis-Gual:2014ewa,Sanchis-Gual:2015sxa,Sanchis-Gual:2016jst,Barranco:2017aes,Huang:2017nho,Sporea:2019iwk}. The study of resonances of various fields around black holes has stimulated our interest in the evolution of resonances in theories of extra dimensions and braneworld.

History of extra dimensions and braneworld dates back to the last century. In the 1920s, Kaluza and Klein (KK) proposed a five-dimensional spacetime theory to unify electromagnetic and gravitational interactions~\cite{kaluza:1921un,Klein:1926tv}. Subsequently, extra-dimensional theories remained silent for more than 70 years. Until the 1990s, to solve the huge hierarchy between the Planck and weak scales, some braneworld models were proposed. Two of them have attracted the attention of many researchers. One is the large extra dimension model proposed by Arkani-Hamed, Dimopoulos, and Dvali~\cite{ArkaniHamed:1998rs}, the other is the warped extra dimension model proposed by Randall and Sundrum~\cite{Randall:1999ee}. The size of extra dimensions is finite in these braneworld models. Subsequently, Antoniadis $et~al.$ embedded the braneworld model with large extra dimensions into the string theory~\cite{Antoniadis:1998ig}. A great development was attained in Ref.~\cite{Randall:1999vf}, which shows that even the extra dimension is infinite, the four-dimensional gravity can also be recovered on the brane. After that, extra-dimensional theories have attracted a lot of attention \cite{Goldberger:1999uk,Gremm:1999pj,DeWolfe:1999cp,Csaki:2000fc,Bazeia:2008zx,Charmousis:2001hg,
Arias:2002ew,Barcelo:2003wq,Bazeia:2004dh,CastilloFelisola:2004eg,Kanno:2004nr,BarbosaCendejas:2005kn,
Koerber:2008rx,BarbosaCendejas:2007vp,Johnson:2008kc,Liu:2011wi,Chumbes:2011zt,Andrianov:2012ae,
Kulaxizi:2014yxa,Dutra:2014xla,Chakraborty:2015zxc,Karam:2018squ}.

In this paper, we focus on the evolution of the scalar KK resonances on the thick brane, which is generated dynamically by a background scalar field. In a braneworld model, to recover the physics in our four-dimensional spacetime, the zero modes of various fields should be localized on the brane. But in addition to zero modes, there are massive KK modes which might propagate into extra dimensions. KK resonances are a specific class of massive KK modes in braneworld models. Usually, for a volcano-like potential, although massive KK modes cannot be localized on the brane, but KK resonances could be quasi-localized on the brane \cite{Liu:2009ve}. In previous literatures, resonances of various fields on thick branes have been studied \cite{Liu:2009ve,Almeida:2009jc,Cruz:2013uwa,Xu:2014jda,Csaki:2000pp,Zhang:2016ksq,Sui:2020fty,Tan:2020sys,Chen:2020zzs}. As far as we know, evolution of KK resonances of various fields in thick brane models has not been studied. The dynamics and the final state of such long-lived modes are still unclear. In the thin brane models, scattering of KK gravitons in the Randall-Sundrum-II model has been considered \cite{Seahra:2005wk,Seahra:2005iq}. It was proved that the brane possesses a set of discrete quasi-normal modes that appear as scattering resonance, and the KK modes of graviton have a very short lifetime on the brane. Can KK modes exist in a thick brane for a long time like the long-lived resonance modes around a black hole? To answer this question, we take the scalar field as a simple example to study the evolution of scalar KK modes numerically. We will show the evolution behaviors of the scalar KK modes and obtain their half-life time on a brane. Further, we will analyze the feasibility of the KK resonances as one of the dark matter candidates. Although our research is still crude, it provides a stepping stone to further  investigation of KK resonances of various fields on a thick brane.

The layout of the remaining part of this paper is as follows. In Sec.~\ref{BRANE WORLD MODEL}, we construct a thick brane solution  in five-dimensional spacetime as the background of a test scalar field evolution. In Sec.~\ref{Scalar field resonances}, the scalar field is evolved with the maximally dissipative boundary condition. Both resonances and nonresonances are used as initial data and their evolution behaviors are compared. Finally, the conclusions and discussions are presented in Sec.~\ref{Conclusion}.

%\section{BRANE WORLD MODEL IN GENERAL RELATIVITY}
\section{Braneworld model in general relativity}
\label{BRANE WORLD MODEL}
Firstly, we consider the thick brane in five-dimensional spacetime. For the simplest case of general relativity with a canonical scalar field, the action is given by
\begin{eqnarray}
  S=\int d^5x\sqrt{-g}\left(\frac{1}{2\kappa^{2}_{5}}R-\frac{1}{2}g^{MN}\partial_{M}
  \phi\partial_{N}\phi-V(\phi)\right),\label{action}
\end{eqnarray}
where $\kappa_{5}$ is the five-dimensional gravitational constant. We set $\kappa_{5}=1$ in this paper for convenience. Besides, we consider the metric of the static flat brane as
\begin{equation}
   ds^2=e^{2A(y)}\eta_{\mu\nu}dx^\mu dx^\nu+dy^2.
   \label{metric}
\end{equation}
Here, $e^{2A(y)}$ is the wrap factor and $\eta_{\mu\nu}=\text{diag}(-1,1,1,1)$ is the four-dimensional Minkowski metric. In this paper, capital Latin letters $M,N,\cdots= 0,1,2,3,5$ label the five-dimensional indices, while Greek letters $\mu,\nu,\cdots= 0,1,2,3$ label the four-dimensional ones. The dynamical field equations are
\begin{eqnarray}
R_{MN}-\frac{1}{2}Rg_{MN}=T_{MN},\label{field equation}\\
g^{MN}\nabla_{M}\nabla_{N}\phi=\frac{\partial V(\phi)}{\partial\phi}.\label{motion equation}
\end{eqnarray}
By substituting the metric~(\ref{metric}) into Eqs.~(\ref{field equation}) and (\ref{motion equation}), we can obtain the explicit equations of motion
\begin{eqnarray}
6A'^2 +3A''&=&-V-\frac{1}{2}\phi'^2,  \label{EoMs1}\\
6A'^2&=&\frac{1}{2}\phi'^2-V,  \label{EoMs2}\\
\phi{''}+4A'\phi'&=&\frac{\partial V}{\partial\phi},  \label{EoMsphi}
\end{eqnarray}
where prime denotes the derivative with respect to the extra dimensional coordinate $y$. Note that only two of the above equations are independent, but we need to solve three functions: $A(y),~\phi(y)$, and $V(\phi)$. So we need to give one of the three functions to solve these equations. The warp factor is
\begin{eqnarray}
A(y)=\ln\Big[\tanh\big(k(y+b)\big)-\tanh \big(k(y-b)\big)\Big],\label{warp factor}
\end{eqnarray}
where the parameter $b$ has length dimension one and the parameter $k$ has mass dimension one. From Eqs.~(\ref{EoMs1}),~(\ref{EoMs2}),~(\ref{EoMsphi}), and (\ref{warp factor}) we get the solution
\begin{eqnarray}
\phi(y)&=&-i \sqrt{3} \text{sech}(b k)\Big[\cosh (2 b k) \text{F}\big(i k y;\tanh ^2(b k)+1\big)\nn\\
&&-2 \sinh ^2(b k) {\rm \Pi} \big(\text{sech}^2(b k);i k y;\tanh ^2(b k)+1\big)\Big],\label{grscalar}\nn\\ \\
V(y)&=&\frac{3}{4} k^2 \Big[-4 \big(\tanh (k (y-b))+\tanh (k (b+y))\big)^2\nn\\
&&+\text{sech}^2(k (y-b))+\text{sech}^2(k (b+y))\Big],
\end{eqnarray}
where $\text{F}(y,q)$ is the first kind elliptic integral and $\Pi(y,q,p)$ is the third kind elliptic integral. This solution was investigated in Ref.~\cite{Tan:2020sys} in $f(T)$ gravity theory. Based on this braneworld background, we consider a test scalar field and study its evolution. We investigate its dynamic behavior numerically, and clarify whether their KK modes can exist on the brane for a long time.

\section{Scalar field resonances and its evolution in braneworld}
\label{Scalar field resonances}
In this section we will consider the evolution of a scalar field in the thick brane given above.  Here we consider a free massless test scalar field $\psi(x^{M})$. Notice that the scalar field $\psi(x^{M})$ here is not the background scalar field $\phi(y)$ that generates the thick brane. The equation of motion for the test field is the Klein-Gordon equation
\begin{equation}
\square^{(5)} \psi=\frac{1}{\sqrt{-g}}\partial_{M}(\sqrt{-g}g^{MN}\partial_{N}\psi)=0.\label{KG equation}
\end{equation}
With the coordinate transformation $dz=e^{-A}dy$, the metric~(\ref{metric}) becomes
\begin{equation}
   ds^2=e^{2A(z)}\eta_{\mu\nu}dx^\mu dx^\nu+e^{2A(z)}dz^2,
   \label{conformalmetric}
\end{equation}
and Eq.~(\ref{KG equation}) can be written as
\begin{equation}
\left[\partial^{2}_{z}+3(\partial_{z}A)\partial_{z}+\eta^{\mu\nu}\partial_{\mu}\partial_{\nu}\right]\psi=0.\label{KGequation1}
\end{equation}
Then, we introduce the following decomposition
\begin{equation}
\psi(x^{M})=e^{-\frac{3}{2}A(z)}\Phi(t,z)\Xi(x^{i}).\label{decomposition1}
\end{equation}
Substituting the above decomposition~(\ref{decomposition1}) into Eq.~(\ref{KGequation1}), we get the following equation
\begin{equation}
-\partial_{t}^{2}\Phi+\partial_{z}^{2}\Phi-U(z)\Phi-a^{2}\Phi=0,\label{evolutionequation}
\end{equation}
where $a$ is a constant from the separation of variables. The effective potential $U(z)$ has the following form:
\begin{equation}
U(z)=\frac{3}{2}\partial_{z}^{2}A+\frac{9}{4}(\partial_{z}A)^{2},\label{effectivepotential}
\end{equation}
or in the $y$ coordinate equivalently
\begin{equation}
U(z(y))=\frac{3}{2}\partial_{y}^{2}A(y)e^{2A(y)}+\frac{15}{4}\left(\partial_{y}A(y)e^{A(y)}\right)^{2}.\label{effectivepotentialy}
\end{equation}
The function $\Phi(t,z)$ can be further decomposed into oscillating modes as
\begin{equation}
\Phi(t,z)=e^{i\omega t}u(z).\label{decomposition2}
\end{equation}
Substituting the above decomposition in Eq.~(\ref{evolutionequation}) yields
\begin{equation}
-\partial_{z}^{2}u(z)+U(z)u(z)=m^{2}u(z),\label{Schrodingerlikeequation}
\end{equation}
where $m^2=\omega^2-a^2$ is the mass of the KK modes. We shall see below that, the effective potential $U(z)$ is a volcano-like potential with a double-well. When $z\rightarrow \pm\infty$, $U(\pm\infty)\rightarrow 0_{+}$. Thus, all KK modes with $m^{2}>0$ are free states. Only the modes with $m^{2}\leq0$ could be bound states. One can show that there is no mode with $m^{2}<0$. The Schr\"odinger-like equation (\ref{Schrodingerlikeequation}) can be factorized as the supersymmetric quantum mechanics form
\begin{equation}
\left(\partial_{z}+\frac{3}{2}\partial_{z}A(z)\right)\left(-\partial_{z}+
\frac{3}{2}\partial_{z}A(z)\right)u(z)=m^{2}u(z).
\end{equation}
Using this supersymmetric quantum mechanics form with the conditions that $A(z)$ is a real function and the extra dimension $z$ is noncompact, it can be shown that there is no tachyon mode with $m^{2}<0$~\cite{Yang:2017puy,Wan:2020smy}. On the other hand, the solution of the zero mode with $m^{2}=0$ is
\begin{equation}
u_{0}(z)\propto e^{\frac{3}{2}A(z)}.
\end{equation}
Obviously, for the warp factor (\ref{warp factor}), the zero mode $u_{0}(z)$ is bound on the brane and it has no node along the extra dimension. According to the node theorem, the zero mode must be the ground state, which ensures that $m^2\geq0$.
Solving Eq.~(\ref{Schrodingerlikeequation}) we could get a series resonant modes, which can be treated as the initial data for the scalar field. The evolution is dominated by Eq.~(\ref{evolutionequation}).

\subsection{Scalar field resonances}
\begin{figure*}
 \centering
\subfigure[~The effective potential (\ref{GReffectiveP})]{\label{figGReffectiveP}
\includegraphics[width=0.4\textwidth]{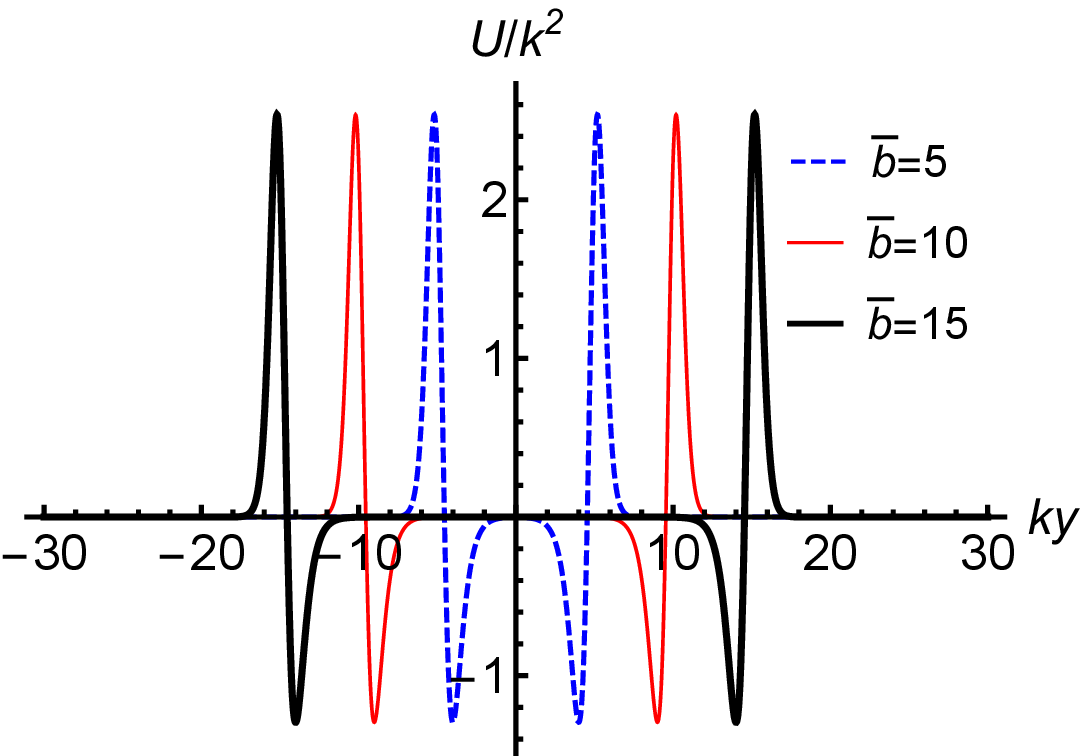}}
\subfigure[~The relative probability with~$\bar{b}=5$]{\label{grP1}
\includegraphics[width=0.4\textwidth]{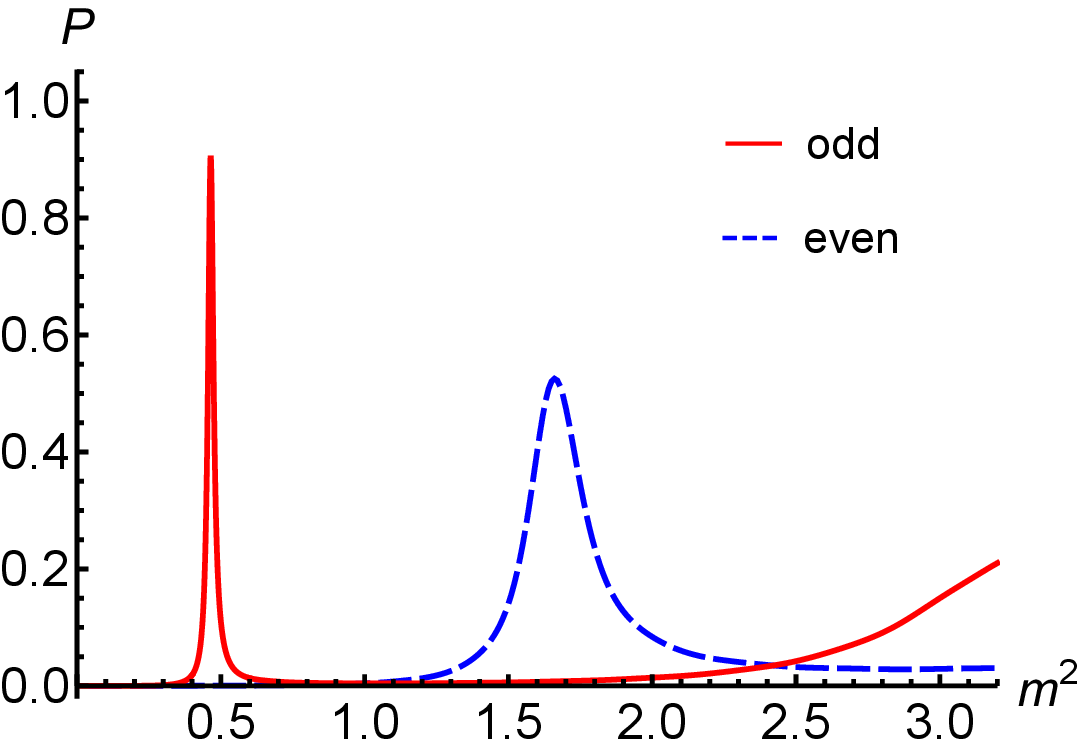}}
\subfigure[~The relative probability with~$\bar{b}=10$]{\label{grP2}
\includegraphics[width=0.4\textwidth]{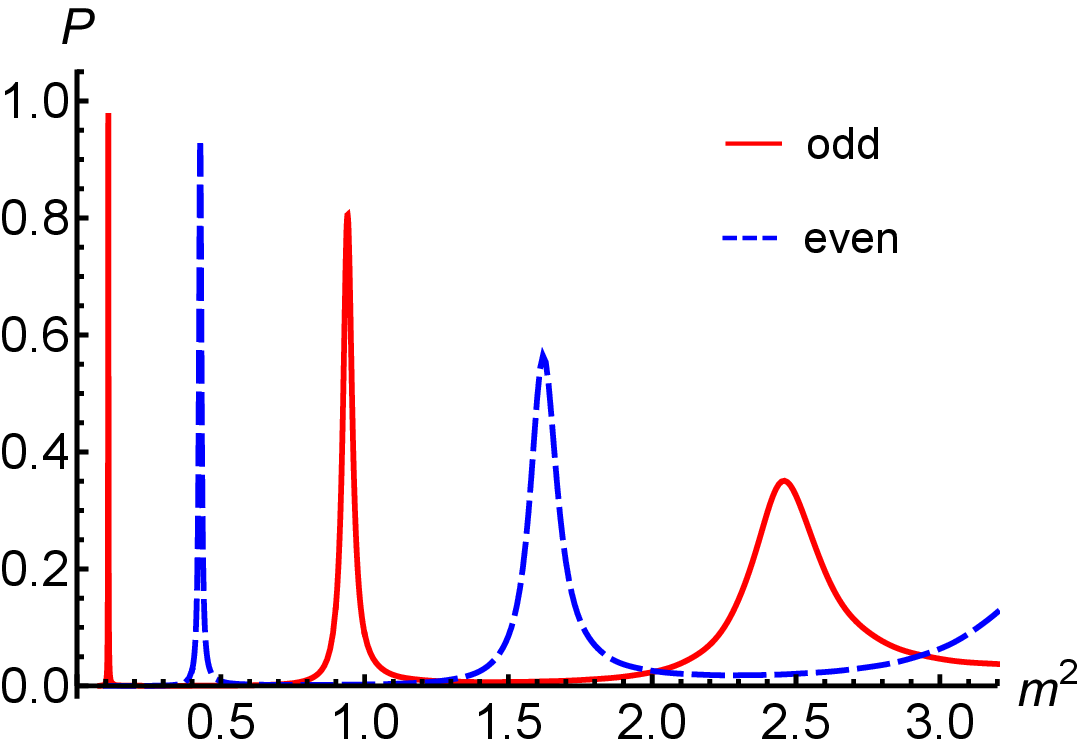}}
\subfigure[~The relative probability with~$\bar{b}=15$]{\label{grP3}
\includegraphics[width=0.4\textwidth]{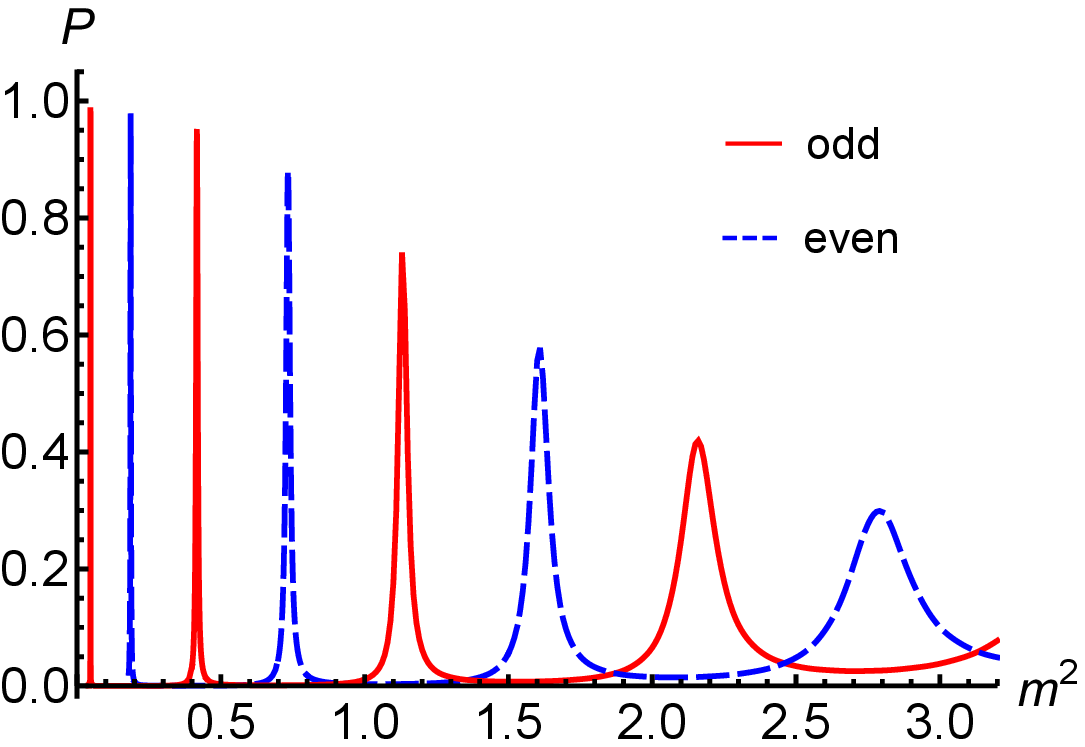}}
\caption{The shapes of the effective potential (\ref{GReffectiveP}) for different parameters $\bar{b}$. And the influence of the parameter $\bar{b}$ on the relative probability $P$ for the odd-parity (red lines) and even-parity (blue dashed lines) massive KK modes. These figures are from Ref.~\cite{Tan:2020sys}.}\label{model1P}
\end{figure*}

In this part, we give a brief review on how to solve the KK scalar resonances. Substituting the warp factor~(\ref{warp factor}) into Eq.~(\ref{effectivepotentialy}), the effective potential in the coordinate $y$ is given by~\cite{Tan:2020sys}
\begin{eqnarray}
U(z(y))&=&-\frac{3}{8}k^2 \text{sech}^2\big(k (b-y)\big) \text{sech}^2\big(k (b+y)\big)\nn\\
 &&\times \Big(\tanh \big(k (b-y)\big)
  +\tanh \big(k (b+y)\big)\Big)^{2} \nn\\
  &&\times\Big(-5\cosh (4 k y)+2 \cosh \big(2 k (b-y)\big)\nn\\
 &&+2 \cosh \big(2 k (b+y)\big)+9\Big)
 \label{GReffectiveP}.
\end{eqnarray}
Plots of the above effective potential are shown in Fig.~\ref{figGReffectiveP}. For convenience, we define the dimensionless parameters $\bar{b}=kb$ and $\bar{m}=m/k$. We can see that the width of the effective potential increases with $\bar{b}$. The resonant modes can be studied by the relative probability method which was proposed in Ref.~\cite{Liu:2009ve}. The relative probability is defined as
\begin{eqnarray}
P(m^{2})=\frac{\int^{z_{b}}_{-z_{b}}|u(z)|^{2}dz}
{\int^{z_{max}}_{-z_{max}}|u(z)|^{2}dz},\label{relative probability}
\end{eqnarray}
where $u(z)$ is solved from Eq.~(\ref{Schrodingerlikeequation}), $z_b$ is approximately the width of the brane, and $z_{max}$ is a much larger width than $z_b$, and usually set to $10z_b$. If the relative probability has a peak with full width at half maximum around $m=m_{n}$, then there is a resonance with mass $m_{n}$. In this way, the modes whose amplitudes in the quasi-well are much larger than those outside the quasi-well can be found. These modes will remain on the brane for a longer time. Note that, the wave functions can be even or odd because the potential is symmetric. Hence, the following boundary conditions can be used to solve Eq.~(\ref{Schrodingerlikeequation}) numerically:
\begin{subequations}
\begin{eqnarray}
\label{even}
u_{\rm{even}}(0)\!\!&=&\!\!1, ~~~\partial_{z}u_{\rm{even}}(0)=0;\\
\label{odd}
u_{\rm{odd}}(0)\!\!&=&\!\!0, ~~~~\partial_{z}u_{\rm{odd}}(0)=1,
\end{eqnarray}\label{EvenOddConditions}
\end{subequations}
where $u_{\rm{even}}$ denotes the even modes of $u(z)$ and $u_{\rm{odd}}$ denotes odd modes of $u(z)$. Substituting the effective potential (\ref{GReffectiveP}) into the Schr\"odinger-like equation~(\ref{Schrodingerlikeequation}), we can solve the solution of $u(z)$ numerically for a given mass $m$. Then the relative probability $P(m^2)$ can be obtained. The relative probability $P(m^2)$ of scalar resonances for $\bar{b}=5,10,15$ are shown in Figs.~\ref{grP1}, \ref{grP2}, \ref{grP3}, respectively. The specific parameters of these resonances are listed in Tab.~\ref{tab1}. It can be seen that the number of scalar resonances and their peak value increase with $\bar{b}$. Usually, the larger peak value means smaller full width at half maximum, and longer lifetime. We will see that in the next subsection.

\begin{table*}[htbp]
\begin{center}
\begin{tabular}{| c |c| c| c| c |}
\hline
$\;\;\bar{b}\;\;$  &
\;\;\;\;\;\;\;\;parity\;\;\;\;\;\;\; &
$\;\;\;\;\;\;\;\;\bar{m}_{n}^{2}\;\;\;\;\;\;\;$ &
$\;\;\;\;\;\;\;\;\bar{m}_{n}\;\;\;\;\;\;\;$ &
$\;\;\;\;\;\;\;\;P\;\;\;\;\;\;\;$\\
\hline
5      &odd   &  0.4649 & 0.6818  & 0.9064 \\
      &even  &  1.6609 & 1.2888  & 0.5253 \\
\hline
      &odd   &  0.1088 & 0.3298  & 0.9794 \\
      &even  &  0.4284 & 0.6545& 0.9408 \\
10      &odd   &  0.9406 & 0.9698  & 0.8058 \\
      &even  &  1.6216 & 1.2734  & 0.5648 \\
      &odd   &  2.4585 & 1.5680  & 0.3508 \\
\hline
      &odd   &  0.0469 & 0.2165  & 0.9892 \\
      &even  &  0.1866 & 0.4320  & 0.9803 \\
      &odd   &  0.4169 & 0.6457  & 0.9516 \\
15      &even  &  0.7337 & 0.8566  & 0.8819 \\
     &odd   &  1.1323 & 1.0641  & 0.7523 \\
      &even  &  1.6078 & 1.2680  & 0.5796 \\
      &odd   &  2.1585 & 1.4692  & 0.4193 \\
      &even  &  2.7892 & 1.6701  & 0.2992 \\
\hline
\end{tabular}
\end{center}
\caption{Resonant mass spectrum $\bar{m}_{n}^{2}$, $\bar{m}_{n}$, and relative probability $P$ for different values of the parameter $\bar{b}$. \label{tab1}}
\end{table*}

\subsection{Evolution of the scalar field resonances}
Treating the scalar resonances as the initial data, we can evolve the scalar field under the evolution equation~(\ref{evolutionequation}). Through the numerical evolution of the scalar field, we can obtain its lifetime on the brane. In this paper, we only consider the case of $a^2=0$, which means that scalar KK particles travel along the extra dimension at the speed of light at infinity. And we impose the maximally dissipative boundary condition: $\partial_{n}\Phi=\partial_{t}\Phi$~\cite{Megevand:2007uy}, where $n$ is the outward unit normal vector to the boundary. Equation~(\ref{evolutionequation}) is solved numerically using fourth-order finite differences in space, and evolving in time using a method of lines with a third-order Runge-Kutta integrator.

In order to more intuitively display the evolution of the scalar field, we define the conserved energy of the scalar field~\cite{Pavlidou:2000cs}
\begin{equation}
E=\int_{-\infty}^{\infty}\rho_{E}dz,  \label{conserved energy}
\end{equation}
where
\begin{equation}
\rho_{E}=\frac{1}{2}\left((\partial_{t}\Phi)^2+ \left(-\frac{3}{2}\partial_{z}A(z)\Phi+\partial_{z}\Phi\right)^2 \right).  \label{energy density}
\end{equation}

\begin{figure*}
\centering
\subfigure[~$E(t)$ with $\bar{b}=5$]{\label{figenergy5}
\includegraphics[width=0.4\textwidth]{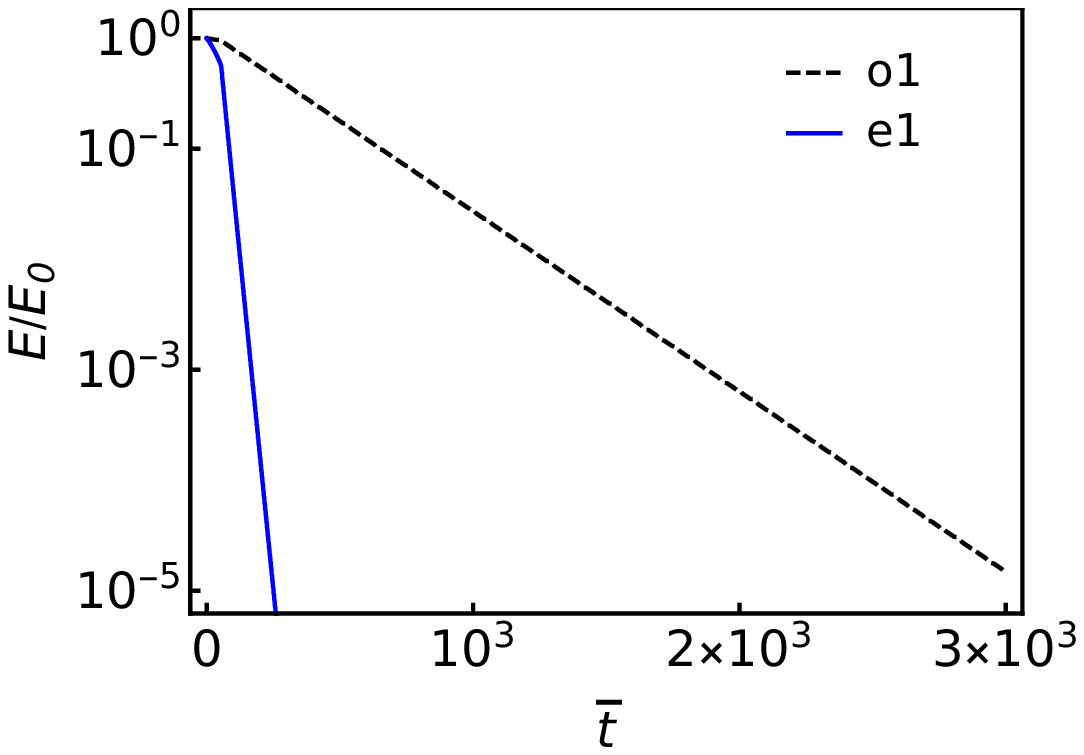}}
\subfigure[~$E(t)$ with $\bar{b}=10$]{\label{figenergy10}
\includegraphics[width=0.4\textwidth]{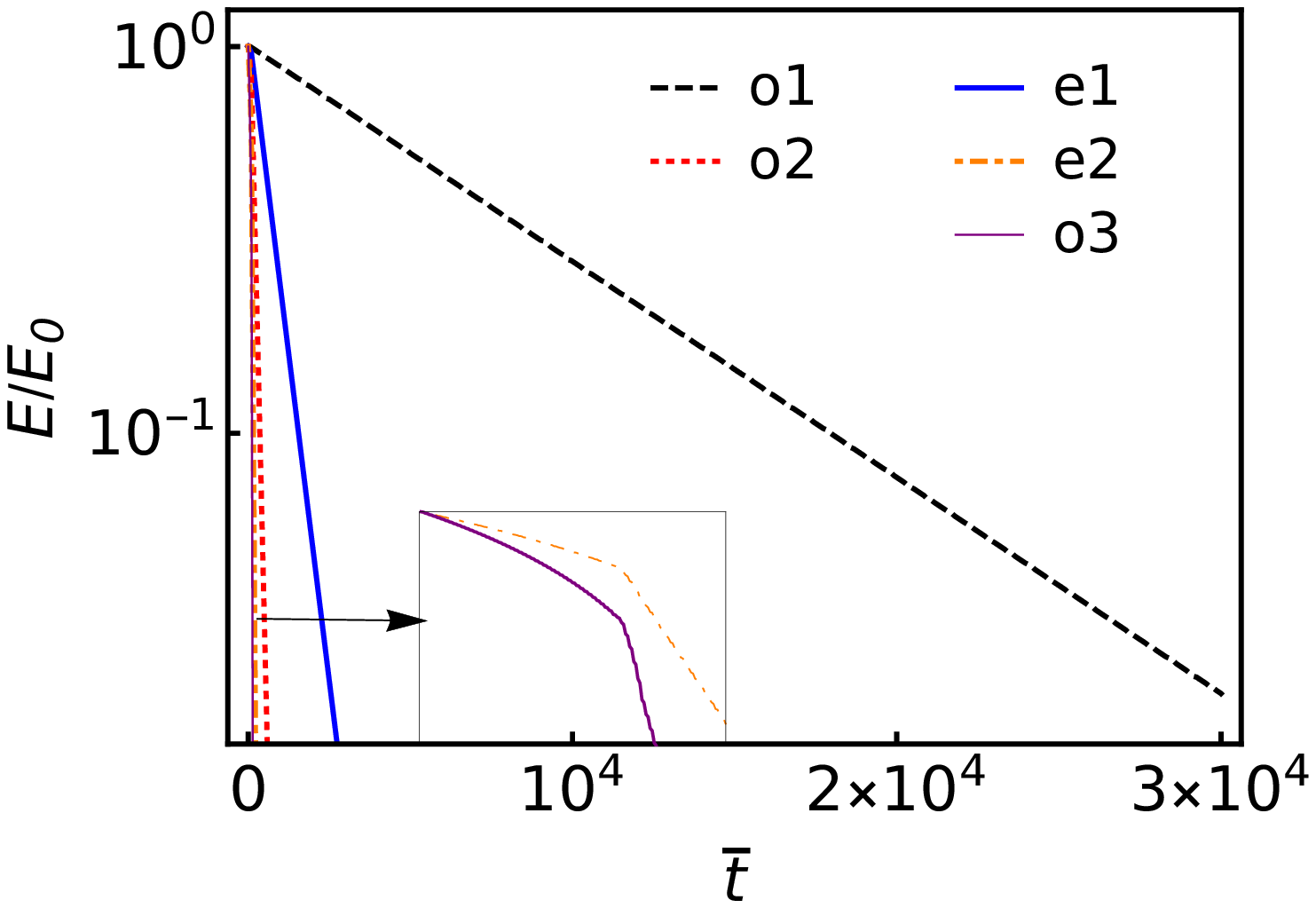}}
\subfigure[~$E(t)$ with $\bar{b}=15$]{\label{figenergy15}
\includegraphics[width=0.4\textwidth]{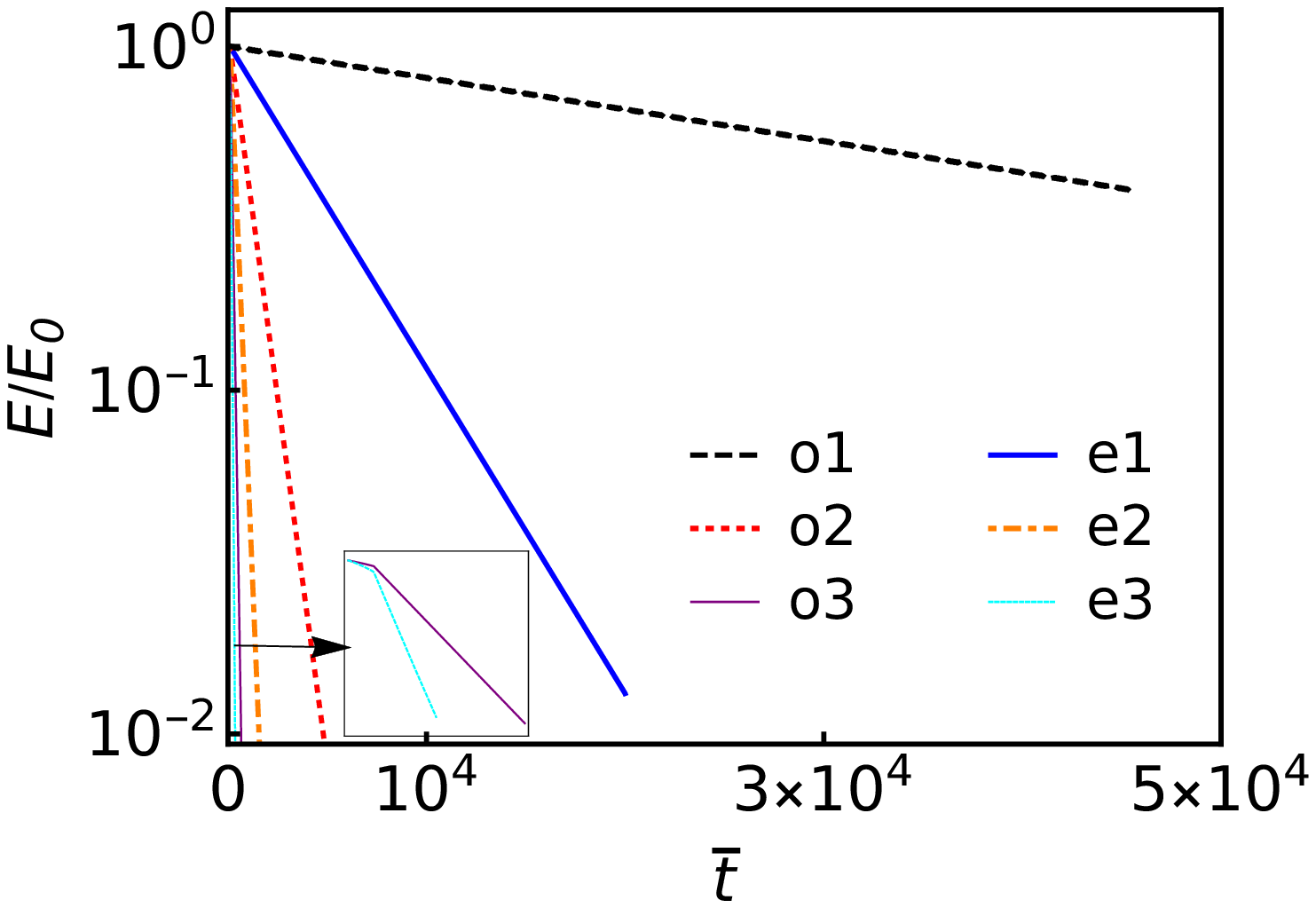}}
\subfigure[$\bar{b}-\bar{t}_{1/2}$]{\label{halflifebfig}
\includegraphics[width=0.4\textwidth]{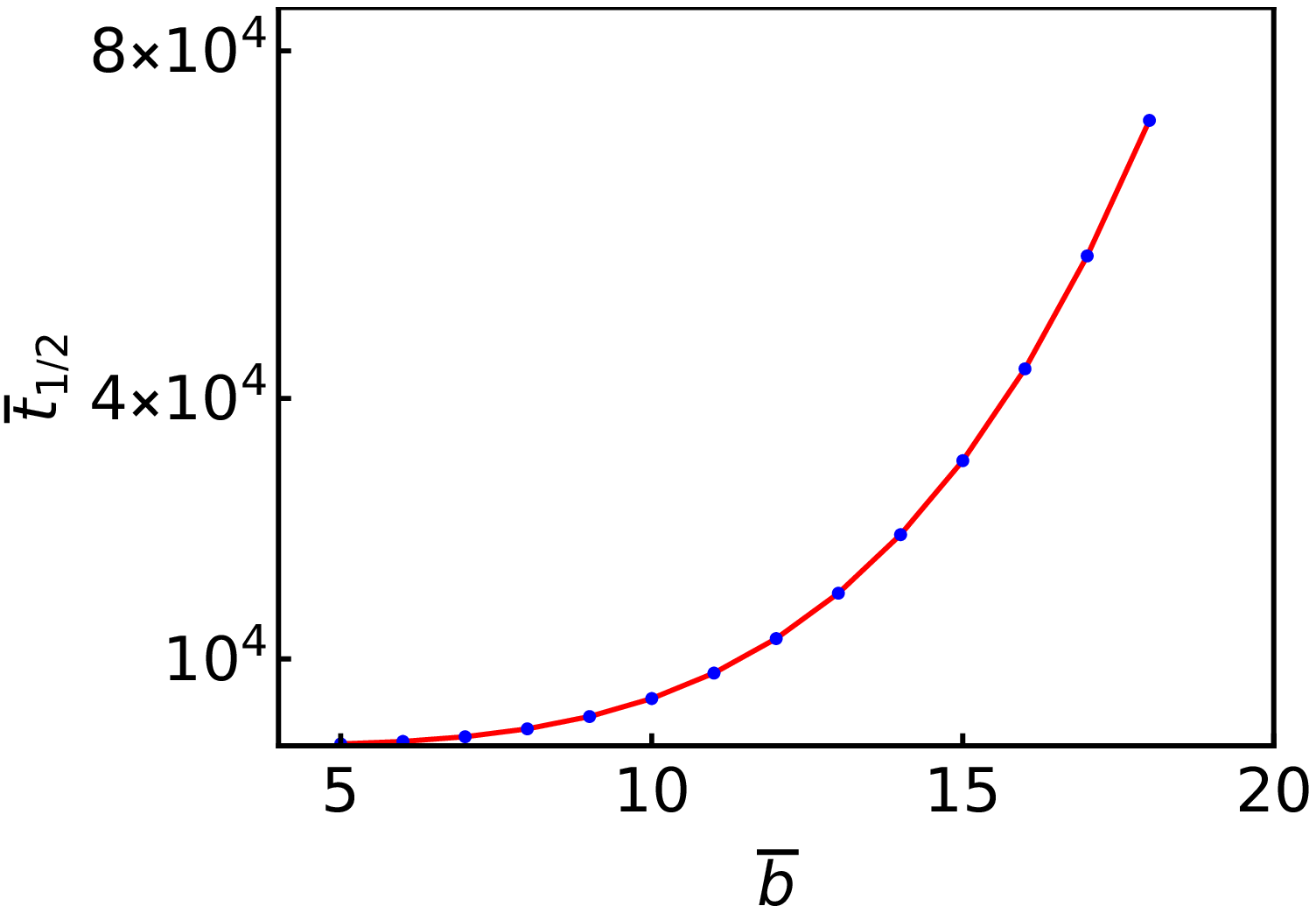}}
\caption{Figures.~\ref{figenergy5}, \ref{figenergy10}, \ref{figenergy15} illustrate the energy of the scalar field vs. time for the evolution of the resonant modes for different values of the parameter $\bar{b}$. Here o1 represents the first odd-parity resonance, e1 represents the first even-parity resonance, and so on. Figure.~\ref{halflifebfig} depicts the relation between the half-life of the first scalar resonance and the parameter $\bar{b}$.}\label{energyfig}
\end{figure*}

Firstly, we consider the evolution of resonances whose parameters are given in Tab.~\ref{tab1}. We integrate the energy density $\rho_{E}$ over $[-z_{max},z_{max}]$, the resulting energy will decay due to energy loses through both left and right boundaries. We plot the evolution of the integrated scalar field energy $E(t)$ in Fig.~\ref{energyfig}. Note that it is plotted on a logarithmic scale. From Figs.~\ref{grP1}, \ref{grP2}, \ref{grP3} we can see that there are several resonances, usually. Among these resonances, the first one will evolve the longest time. Besides, the evolution time increases with the parameter $\bar{b}$. This is because that the larger $\bar{b}$ the more energy of the scalar field is concentrated in the potential well. In addition, we plot half-life time of the first resonance with different values of the parameter $\bar{b}$, which can be seen from Fig~\ref{halflifebfig}. Note that here we have defined the dimensionless time $\bar{t}=kt$. It can be clearly seen that the  half-life time of the resonances increases with the parameter $\bar{b}$.

The energy decay can be fitted as an exponential function:
\begin{eqnarray}
E(t)=E_{0}\text{exp}(-s\bar{t}),\label{energy decay}
\end{eqnarray}
where $s$ is the fitting parameter and $E_{0}$ denotes the initial energy of the KK mode. Some results of the fit are listed in Table~\ref{tab2}. It can be seen that both the scaled mass $\bar{m}_1$ and the fitting parameter $s$ of the first resonance decreases with $\bar{b}$.

\begin{table*}[htbp]
\begin{center}
\begin{tabular}{| c | c| c| c | c |}
\hline
$\;\;\bar{b}\;\;$  &
$\;\;\;\;\;\;\;\;\bar{m}_{1}^{2}\;\;\;\;\;\;\;$ &
$\;\;\;\;\;\;\;\;\bar{m}_{1}\;\;\;\;\;\;\;$ &
$\;\;\;\;\;\;\;\;\;\;s\;\;\;\;\;\;\;\;\;$  &
$\;\;\;\;\;\;\;\;\;t_{1/2}$~(~if~$k=10^{-10}$eV)\;\;\;\;\;\;\;\;\;\\
\hline
5       &  0.4649 & 0.6818  & 3.7527$\times10^{-3}$ & 1.1969$\times10^{-3}$seconds\\
\hline
6        &  0.3177 & 0.5636 & 1.4624$\times10^{-3}$ & 3.0714$\times10^{-3}$seconds\\
\hline
7        &  0.2298 & 0.4794  & 6.9109$\times10^{-4}$ & 6.4993$\times10^{-3}$seconds\\
\hline
8        &  0.1736 & 0.4166  & 3.6373$\times10^{-4}$ & 1.2349$\times10^{-2}$seconds\\
\hline
9        &  0.1356 & 0.3682  & 2.0910$\times10^{-4}$ & 2.1481$\times10^{-2}$seconds\\
\hline
10       &  0.1088 & 0.3298  & 1.2867$\times10^{-4}$ & 3.4908$\times10^{-2}$seconds\\
\hline
11       &  0.0892 & 0.2986  & 8.3435$\times10^{-5}$ & 5.3833$\times10^{-2}$seconds\\
\hline
12      &  0.0744 & 0.2728  & 5.6502$\times10^{-5}$ & 7.9494$\times10^{-2}$seconds\\
\hline
13         &  0.0630 & 0.2510  & 3.9624$\times10^{-5}$ & 1.1334$\times10^{-1}$seconds\\
\hline
14       &  0.0540 & 0.2324  & 2.8609$\times10^{-5}$ & 1.5700$\times10^{-1}$seconds\\
\hline
15        &  0.0469 & 0.2165  & 2.1170$\times10^{-5}$ & 2.1217$\times10^{-1}$seconds\\
\hline
16       &  0.0410 & 0.2026  & 1.6000$\times10^{-5}$ & 2.8072$\times10^{-2}$seconds\\
\hline
17       & 0.0362 & 0.1903  & 1.2316$\times10^{-5}$ & 3.6470$\times10^{-1}$seconds\\
\hline
18       &  0.0322 & 0.1794  & 9.6335$\times10^{-6}$ & 4.6625$\times10^{-1}$seconds\\
\hline
\end{tabular}
\end{center}
\caption{The first resonant mass spectrum $\bar{m}_{1}^{2}$, $\bar{m}_{1}$, fitting parameter $s$, and half-life $t_{1/2}$ for different values of the parameter $\bar{b}$.} \label{tab2}
\end{table*}

\begin{figure*}
\centering
\subfigure[~$\bar{z}_{\text{ext}}=3$]{\label{figa3o1}
\includegraphics[width=0.4\textwidth]{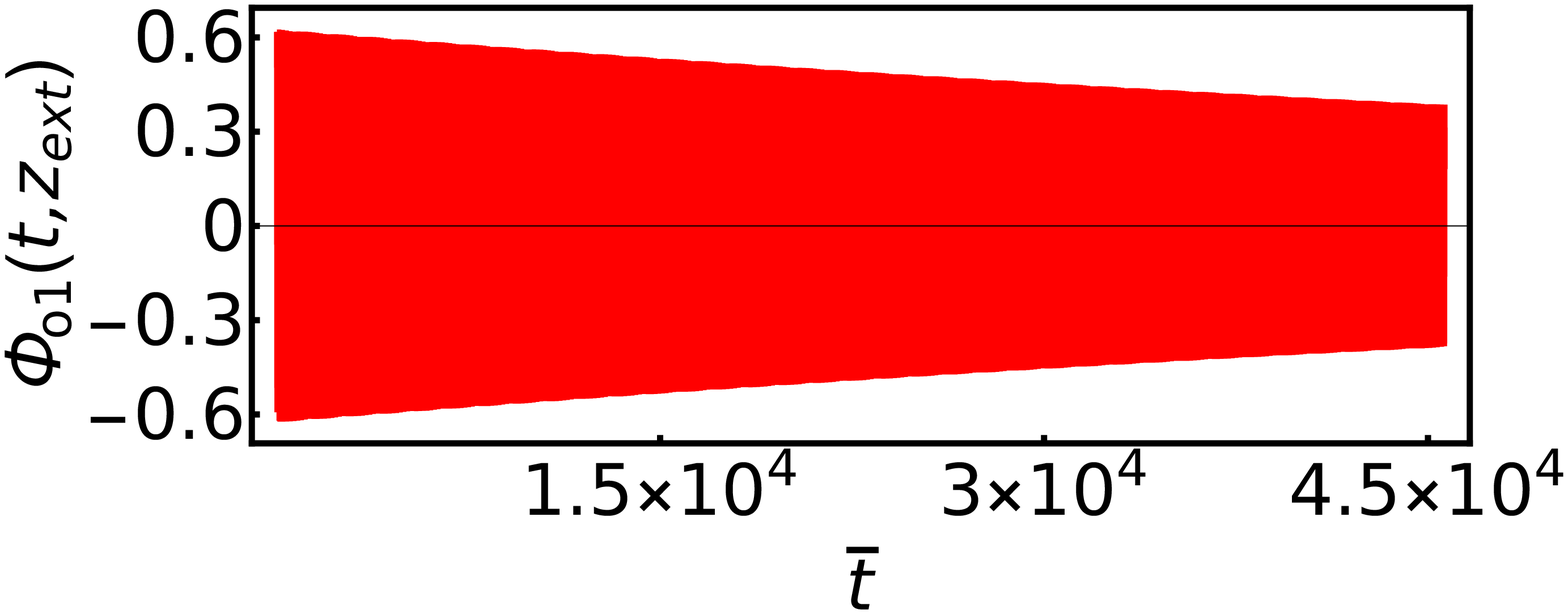}}
\subfigure[~$\bar{z}_{\text{ext}}=3$]{\label{figa10o1}
\includegraphics[width=0.4\textwidth]{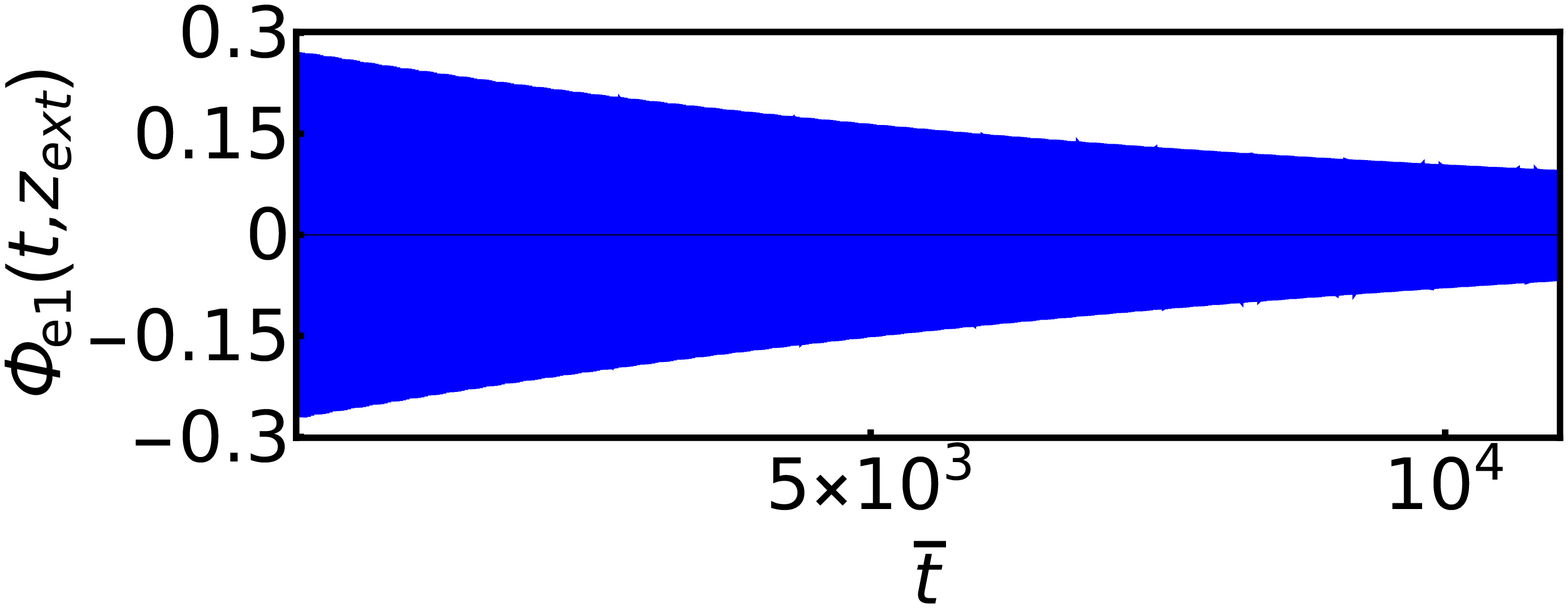}}
\subfigure[~$\bar{z}_{\text{ext}}=30$]{\label{figa10e1}
\includegraphics[width=0.4\textwidth]{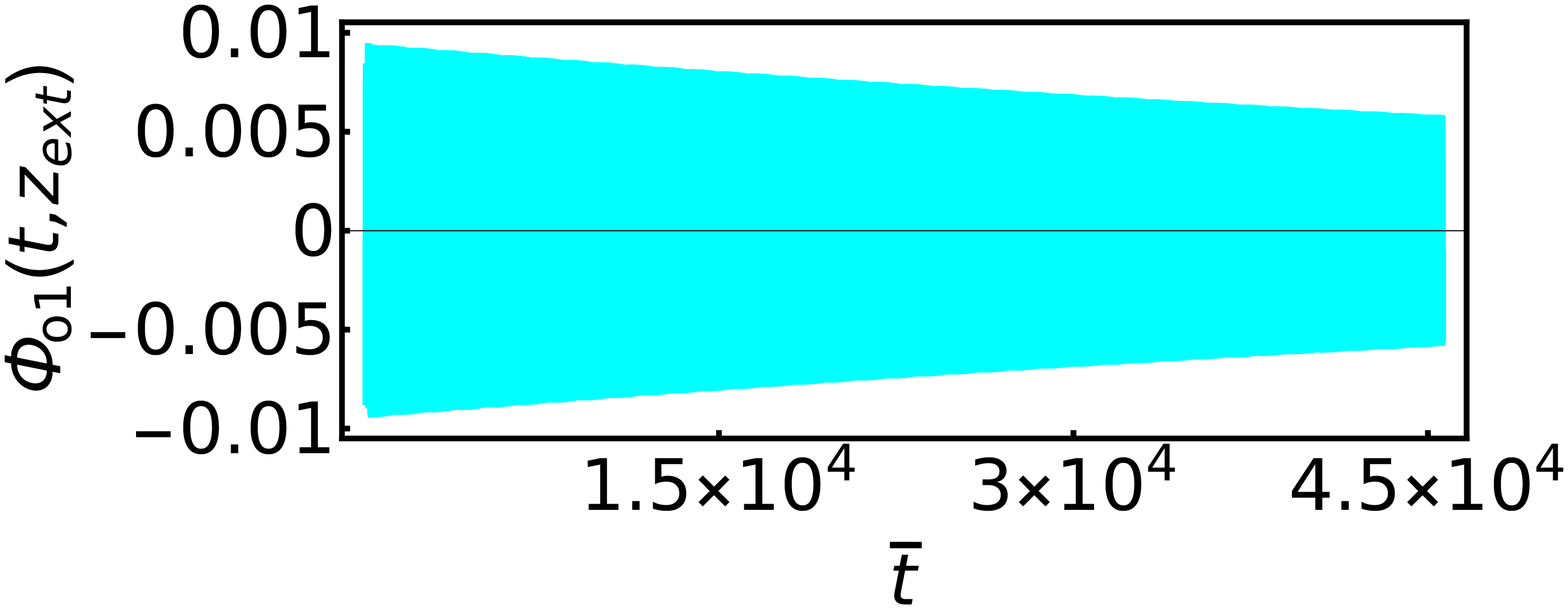}}
\subfigure[~$\bar{z}_{\text{ext}}=30$]{\label{figa3e1}
\includegraphics[width=0.4\textwidth]{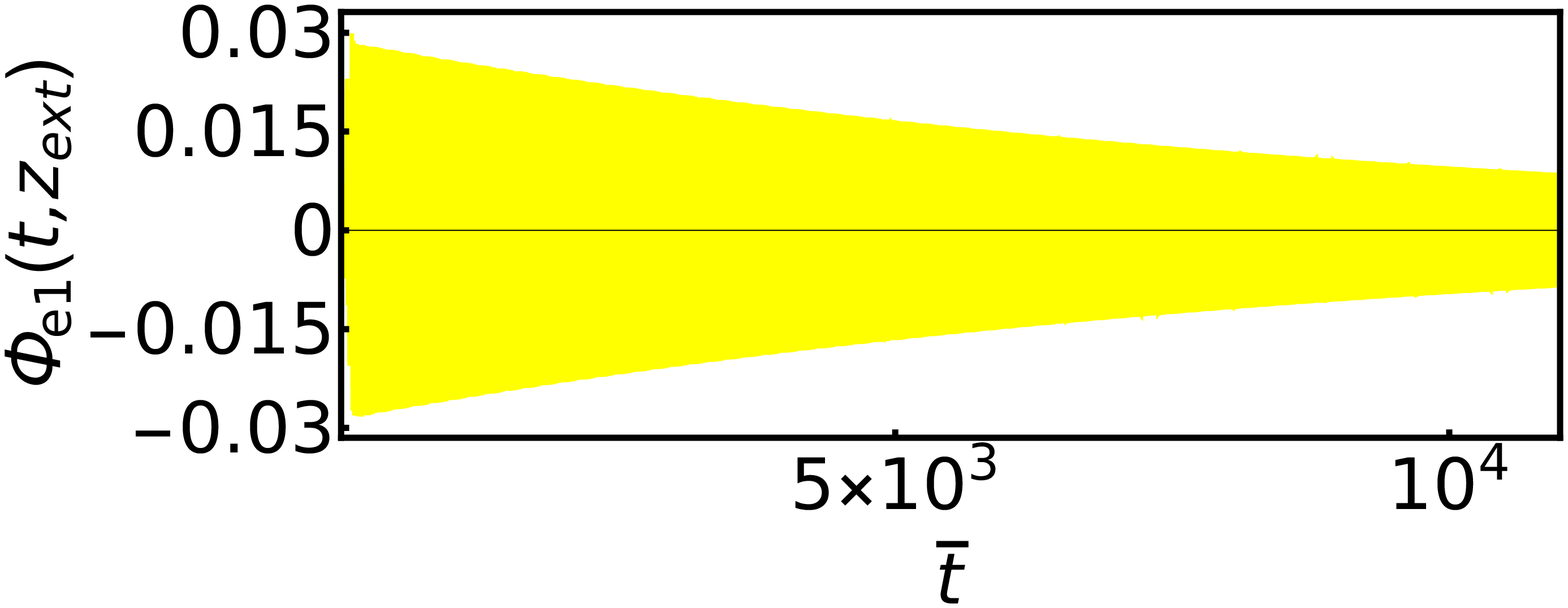}}
\caption{Upper panel: Time evolution of the first odd resonance (left) and the first even resonance (right) at $\bar{z}_{\text{ext}}=3$ for $\bar{b}=15$.
Lower panel: Time evolution of the first odd resonance (left) and the first even resonance (right) at $\bar{z}_{\text{ext}}=30$ for $\bar{b}=15$. Here $\bar{z}_{\text{ext}}=kz_{\text{ext}}.$}\label{extfig}
\end{figure*}

To better show the evolution of the scalar field over time, we analyze the result of the numerical evolution by extracting a time series for the resonance amplitude at a fixed point $z_{\text{ext}}$.
The results are shown in Fig.~\ref{extfig}. It can be seen that the resonance amplitude decreases with evolutionary time, and the amplitude attenuation of the second resonance is significantly faster than that of the first resonance. This also shows that the first resonance will evolve longest. By comparing the amplitude attenuation at different positions of the same resonance, it can be seen that the attenuation rate and the overall shape are basically the same. In other words, at least for the first two resonances, there is no beating effect found in Ref.~\cite{Witek:2012tr} for a black hole system.

\begin{figure*}
\centering
\subfigure[~$\bar{z}_{\text{ext}}=3$]{\label{fignonresonancefourier}
\includegraphics[width=0.4\textwidth]{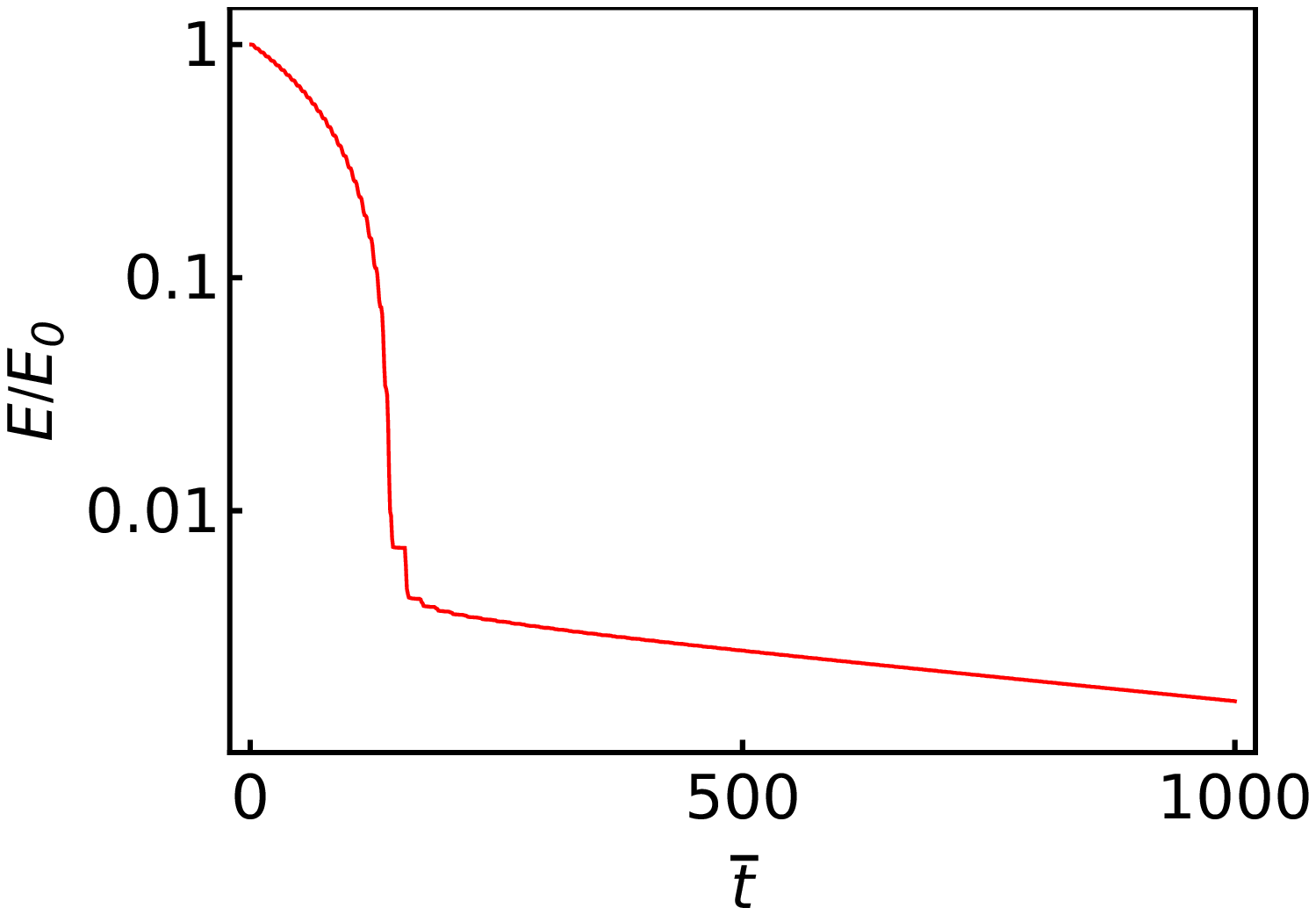}}
\subfigure[~$\bar{z}_{\text{ext}}=30$]{\label{fignonresonancea}
\includegraphics[width=0.4\textwidth]{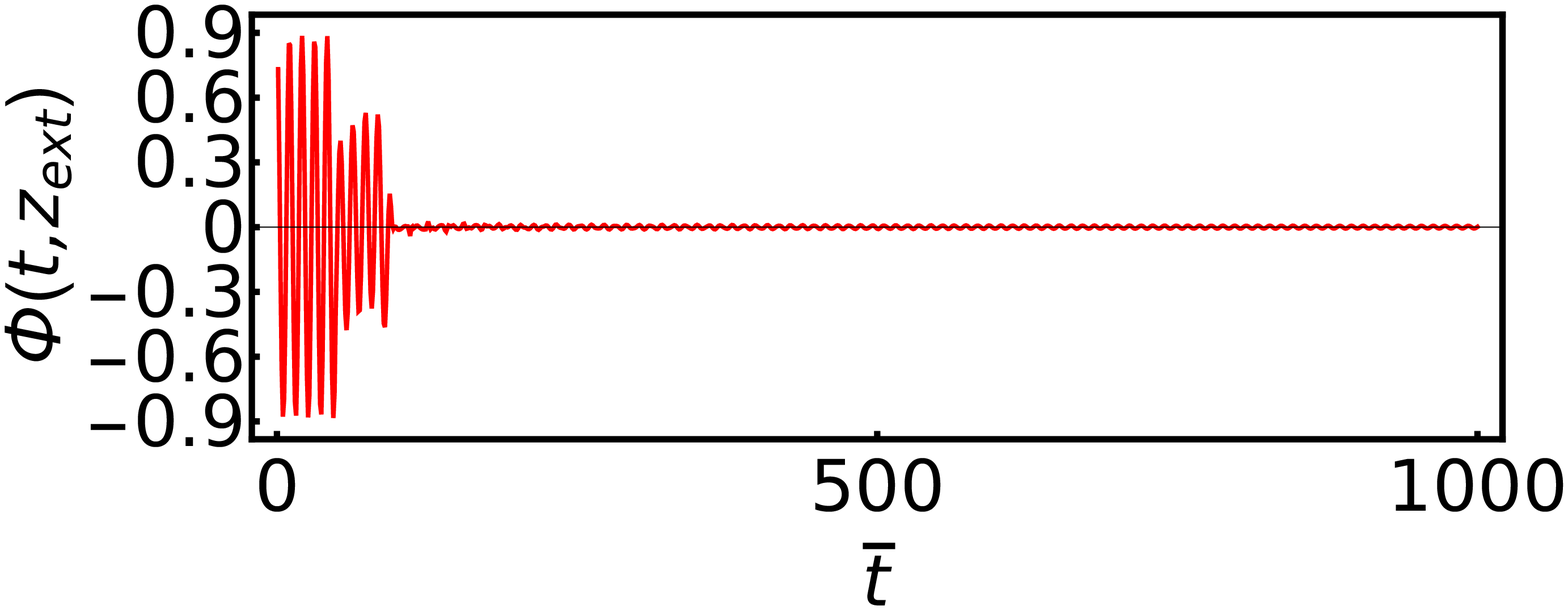}}
\caption{Upper panel: The energy of the scalar field vs. time for the evolution of nonresonance for $\bar{b}=15$. Lower panel: Time evolution of the nonresonance with $\bar{z}_{\text{ext}}=30$ for $\bar{b}=15$. }\label{nonresonancefig}
\end{figure*}

\begin{figure*}
\centering
\subfigure[~$\bar{z}_{\text{ext}}=3$]{\label{fignonresonancefourier}
\includegraphics[width=0.4\textwidth]{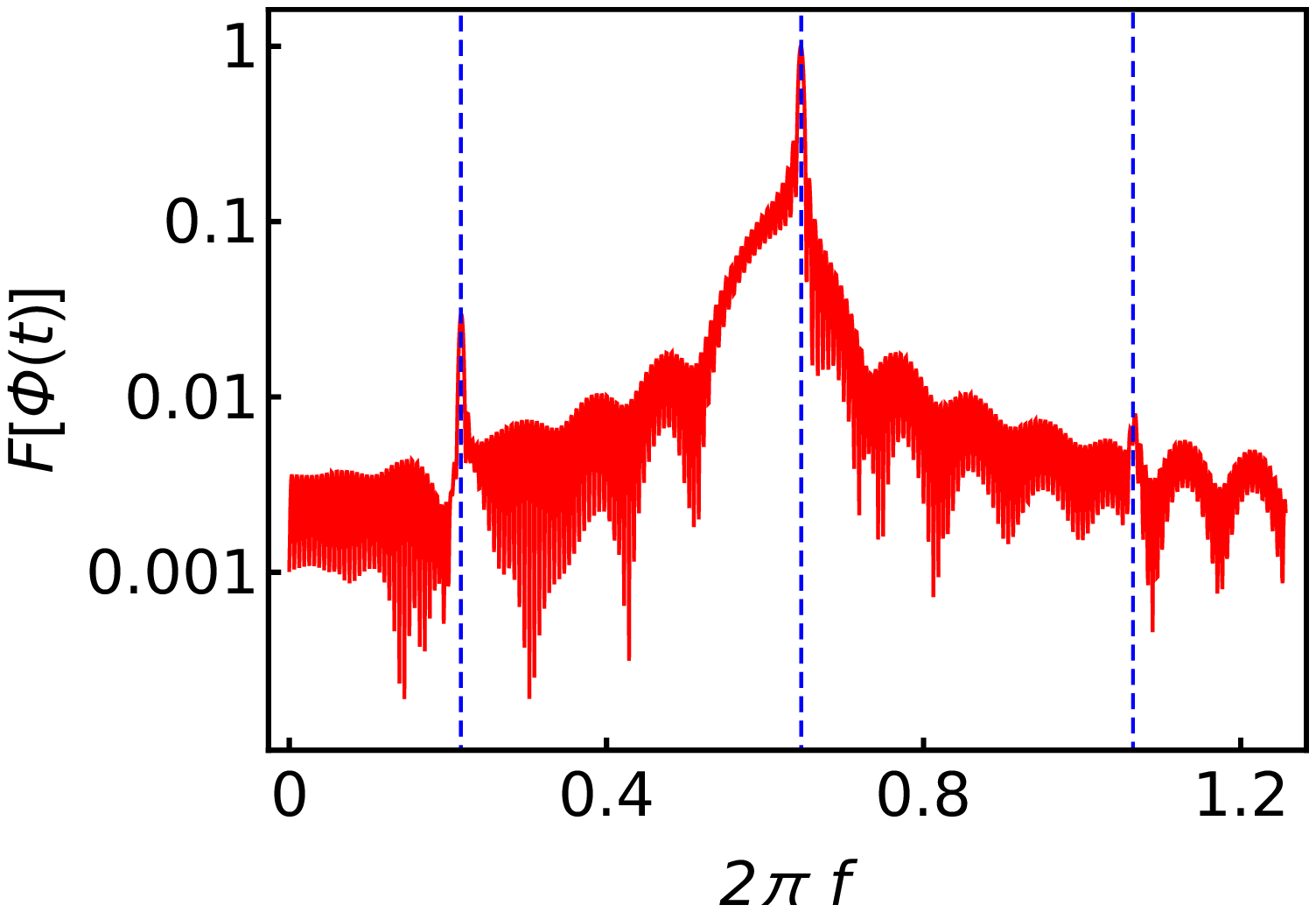}}
\subfigure[~$\bar{z}_{\text{ext}}=3$]{\label{figfourierb15o1}
\includegraphics[width=0.4\textwidth]{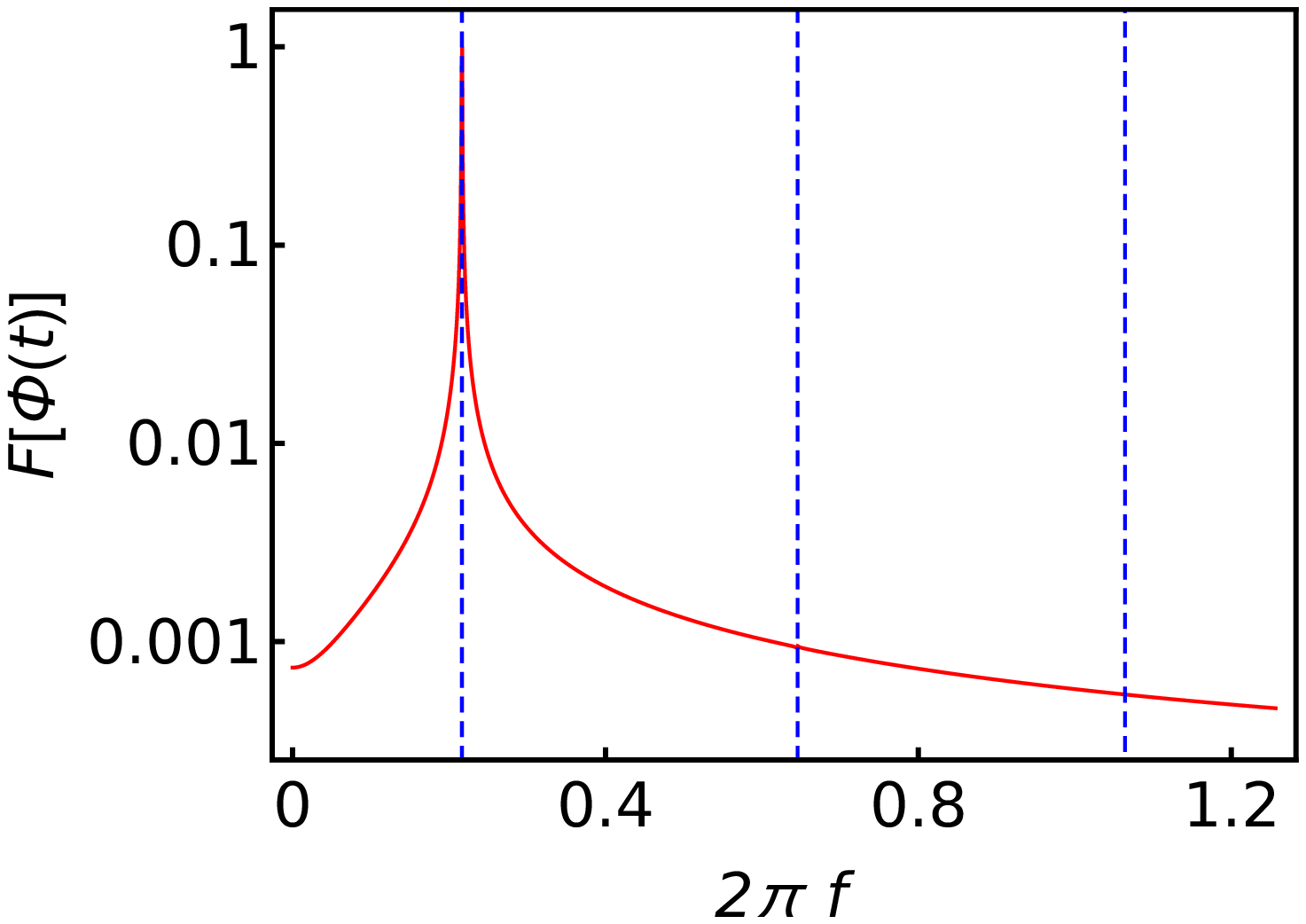}}
\caption{Left panel: Discrete Fourier transform in time vs. frequency for the evolution of the nonresonance with $\bar{m}^2=0.36$ for $\bar{b}=15$. Right panel: Discrete Fourier transform in time vs. frequency for the evolution of the first resonance with $\bar{m}_1^2=0.04687$ for $\bar{b}=15$. The blue dotted lines correspond to the frequencies of the first three odd parity resonances. }\label{fourierfig}
\end{figure*}

As a comparison, we also consider the evolution of the nonresonances.  These results are shown in Fig. \ref{nonresonancefig}. We find that the energy and amplitude of nonresonance decay very fast at early stage, but later they decay like those of resonances. In order to gain a better understanding of the above results, we perform a spectral analysis. We calculate the discrete Fourier transform in time of the scalar field at a fixed point $z=z_{j}$. The explicit expression of the discrete Fourier transform is
\begin{eqnarray}
F[\Phi(t)](f):=|A\sum_{p}\Phi(t_{p},z_{j})\text{exp}(-2\pi ift_{p} )|,\label{Fourier transform}
\end{eqnarray}
where $A$ is normalization constant and $t_{p}$ are the discrete time values. Plots of the Fourier transform for the first resonance and the nonresonance with $\bar{m}^2=0.36$ for $\bar{b}=15$ are shown in Fig.~\ref{fourierfig}. We find that, for the Fourier transform of the resonance, there is only one peak corresponding to the resonance frequency. However, for the Fourier transform of the nonresonance, there are several peaks.  Thus, nonresonances can evolve into combinations of resonances, and from this point of view, resonances seem to play a similar role in the braneworld as the quasi-normal modes in black holes physics, which deserves further investigation.

Finally, we consider the half-life time of the scalar resonances. If the exponential decay~(\ref{energy decay}) is sustained throughout all evolution, then it is easy to know that $\bar{t}_{1/2}=\frac{\ln(2)}{s}$. For $k=10^{-10}\text{eV}$ and $s=10^{-10}$, the half-life time $t_{1/2}$ of the first scalar resonance will reach $10^{4}$ seconds. Admittedly, it is still short compared with the age of our universe. But note that the lifetime of resonances increases with the parameter $\bar{b}$. Thus, for a very large $\bar{b}$, the lifetime of resonance might reach the cosmological time scale.
Here, we can simply estimate the feasibility of resonance as a candidate for dark matter. For the braneworld model in this part, the effective four-dimensional Planck scale $M_{\text{Pl}}$ and the fundamental five-dimensional scale $M_{5}$ have the relation:
\begin{eqnarray}
M_{\text{Pl}}^{2}=\frac{8\bar{b}\coth(2\bar{b})-4}{k}M_{5}^{3}.\label{relation1}
\end{eqnarray}
According to the current experiment of the Large Hadron Collider, the collision energy is 13 TeV but no signal of extra dimensions is seen~\cite{ATLAS:2017tlw}. So the fundamental five-dimensional scale $M_{5}$ should be greater than 13 TeV. Thus, combining Eq. (\ref{relation1}) and the condition $M_{5}\gtrsim 13$ TeV, the constraint on the parameter $k$ is given by
\begin{eqnarray}
k\gtrsim(12\bar{b}\coth(2\bar{b})-6)\times 10^{-17} \text{eV}.
\end{eqnarray}
From Tab. \ref{tab2}, we can know that the fitting parameter $s$ rapidly decreases with $\bar{b}$. In fact, when $\bar{b}$ doubles, the fitting parameter $s$ decreases by an order of magnitude. If we choose $\bar{b}=10^{8}$, thus, $k\geq 10^{-8}$ eV, and $s$ might be $10^{-25}$, and the half-life time $t_{1/2}$ of the first scalar resonance will reach $10^{17}$ seconds. It is the same order of magnitude as the age of our universe ($4.35\times 10^{17}$ seconds). On the other hand, the large $\bar{b}$ means that the first resonance is very light, but they are still hard to spot in the collider. The reason is that the cross section of any process involving the interaction of the zero mode with the light continuum modes is imperceptibly low.
From this perspective, these long-lived resonances with very light mass could be considered as a candidate for dark matter.

\section{Conclusion and discussion}
\label{Conclusion}
In this paper, we investigated the evolution of a free massless scalar field in the thick brane numerically. We find that the resonances decay very slowly compared to the nonresonances and can exist on the brane for a very long time. If the lifetime of these resonances can be long enough as the cosmological time scale, they might be a candidate for dark matter. This provides a new idea for dark matter research.

Firstly, we constructed a five-dimensional thick brane generated by a scalar field. Then, we considered the evolution of a test scalar field in this thick brane background. Through the coordinate transformation and the variable separation, we got the evolution equation~(\ref{evolutionequation}) and the Schr\"odinger-like equation~(\ref{Schrodingerlikeequation}) for the extra dimension profile of the scalar field. The latter gives us the initial data of the massive KK modes, and the former evolves those initial data. Next we solved the Schr\"odinger-like equation~(\ref{Schrodingerlikeequation}) numerically to obtain the initial data of massive KK modes, especially the initial data of the resonances. The results were shown in Fig.~\ref{model1P} and Tab.~\ref{tab1}. We found that the relative probability of the first scalar resonance increases with $\bar{b}$, while the mass $m_{1}$ of the first scalar resonance decreases with $\bar{b}$. Using these KK modes as initial data, we investigated their evolution. Considering their energy decay and  extracting a time series for the resonance amplitude, the evolution of the scalar field was analyzed. The results were shown in Fig.~\ref{energyfig}, Fig.~\ref{extfig}, and Tab.~\ref{tab2}. The energy decay can be described by a decay parameter $s$ which can be obtained by an exponential fit of $E/E_{0}$ as the function of $\bar{t}$. On the other hand, we also considered the evolution of the nonresonances. The energies and amplitudes of nonresonances decay rapidly at early stage, but later they decay like those of the resonances. The behavior of the nonresonance evolution could be treated as a combination of resonances. Finally, we considered the half-life time of the resonance. For a very large $\bar{b}$, the lifetime of ultra-light resonances can reach the cosmological time scale. These indicate that the scalar resonant mode could be one of the candidates for dark matter.

There is much to be improved in this paper. Firstly, the mass and interaction of the test scalar field is not taken into account, which may result in more interesting results. Secondly, other test fields such as Dirac spinor fields and gauge fields, their evolution is also worth investigating. These possibilities deserve further study.

\section*{Acknowledgements}
We are thankful to X.-L.~Du, X.-H.~Zhang, and J.-J.~Wan for useful discussions. This work was supported by the National Key Research and Development Program of China (Grant No. 2020YFC2201503), the National Natural Science Foundation of China (Grants No.~11875151, No.~11705070, No.~12105126, and No.~12047501), the 111 Project under (Grant No. B20063), the Fundamental Research Funds for the Central Universities (Grant No. lzujbky2021-pd08), the China Postdoctoral Science Foundation (Grants No. 2021M701529, and 2021M701531), and ``Lanzhou City's scientific research funding subsidy to Lanzhou University".


\begin{thebibliography}{99}


\bibitem{Planck:2018vyg}
N.~Aghanim \textit{et al.} [Planck],
\emph{Planck 2018 results. VI. Cosmological parameters},
{\emph{Astron. Astrophys.} {\bfseries 641} (2020) A6},
[erratum: Astron. Astrophys. \textbf{652} (2021) C4],
[{{\ttfamily arXiv:1807.06209}}].
%6498 citations counted in INSPIRE as of 13 Dec 2021

\bibitem{Barranco:2011eyw}
J.~Barranco, A.~Bernal, J.~C.~Degollado, A.~Diez-Tejedor, M.~Megevand, M.~Alcubierre, D.~Nunez, and O.~Sarbach,
\emph{Are black holes a serious threat to scalar field dark matter models?}
{\emph{Phys. Rev. D} {\bfseries 84} (2011) 083008},
[{{\ttfamily arXiv:1108.0931}}].


\bibitem{Barranco:2012qs}
J.~Barranco, A.~Bernal, J.~C.~Degollado, A.~Diez-Tejedor, M.~Megevand, M.~Alcubierre, D.~Nunez, and O.~Sarbach,
\emph{Schwarzschild black holes can wear scalar wigs},
{\emph{Phys. Rev. Lett.} {\bfseries 109} (2012) 081102},
[{{\ttfamily arXiv:1207.2153}}].
%69 citations counted in INSPIRE as of 10 Dec 2021

\bibitem{Barranco:2013rua}
J.~Barranco, A.~Bernal, J.~C.~Degollado, A.~Diez-Tejedor, M.~Megevand, M.~Alcubierre, D.~N\'u\~nez, and O.~Sarbach,
\emph{Schwarzschild scalar wigs: spectral analysis and late time behavior},
{\emph{Phys. Rev. D} {\bfseries 89} (2014)  083006},
[{{\ttfamily arXiv:1312.5808 }}].
%27 citations counted in INSPIRE as of 10 Dec 2021


\bibitem{Zhou:2013dra}
X.~N.~Zhou, X.~L.~Du, K.~Yang, and Y.~X.~Liu,
\emph{Dirac dynamical resonance states around Schwarzschild black holes},
{\emph{Phys. Rev. D} {\bfseries 89} (2014)  043006},
[{{\ttfamily arXiv:1308.2863 }}].
%11 citations counted in INSPIRE as of 10 Dec 2021


\bibitem{Gossel:2013cka}
G.~H.~Gossel, J.~C.~Berengut, and V.~V.~Flambaum,
\emph{Resonance scattering and the passage to bound states in the field of near-black-hole objects},
{\emph{Int. J. Mod. Phys. D} {\bfseries 23} (2014) 1450089},
[{{\ttfamily arXiv:1308.6426}}].
%0 citations counted in INSPIRE as of 10 Dec 2021


\bibitem{Decanini:2014bwa}
Y.~D\'ecanini, A.~Folacci, and M.~Ould El Hadj,
\emph{Resonant excitation of black holes by massive bosonic fields and giant ringings},
{\emph{Phys. Rev. D} {\bfseries 89} (2014) 084066},
[{{\ttfamily arXiv:1402.2481}}].
%6 citations counted in INSPIRE as of 10 Dec 2021


\bibitem{Sampaio:2014swa}
M.~O.~P.~Sampaio, C.~Herdeiro, and M.~Wang,
\emph{Marginal scalar and Proca clouds around Reissner-Nordstr\"om black holes},
{\emph{Phys. Rev. D} {\bfseries 90} (2014) 064004},
[{{\ttfamily arXiv:1406.3536}}].
%40 citations counted in INSPIRE as of 10 Dec 2021

\bibitem{Degollado:2014vsa}
J.~C.~Degollado and C.~A.~R.~Herdeiro,
\emph{Wiggly tails: a gravitational wave signature of massive fields around black holes},
{\emph{Phys. Rev. D} {\bfseries 90} (2014) 065019},
[{{\ttfamily arXiv:1408.2589}}].
%32 citations counted in INSPIRE as of 10 Dec 2021


\bibitem{Sanchis-Gual:2014ewa}
N.~Sanchis-Gual, J.~C.~Degollado, P.~J.~Montero, and J.~A.~Font,
\emph{Quasistationary solutions of self-gravitating scalar fields around black holes},
{\emph{Phys. Rev. D} {\bfseries 91} (2015) 043005},
[{{\ttfamily arXiv:1412.8304}}].
%24 citations counted in INSPIRE as of 10 Dec 2021

\bibitem{Sanchis-Gual:2015sxa}
N.~Sanchis-Gual, J.~C.~Degollado, P.~J.~Montero, J.~A.~Font, and V.~Mewes,
\emph{Quasistationary solutions of self-gravitating scalar fields around collapsing stars},
{\emph{Phys. Rev. D} {\bfseries 92} (2015) 083001},
[{{\ttfamily arXiv:1507.08437}}].
%19 citations counted in INSPIRE as of 10 Dec 2021


\bibitem{Sanchis-Gual:2016jst}
N.~Sanchis-Gual, J.~C.~Degollado, P.~Izquierdo, J.~A.~Font, and P.~J.~Montero,
\emph{Quasistationary solutions of scalar fields around accreting black holes},
{\emph{Phys. Rev. D} {\bfseries 94} (2016) 043004},
[{{\ttfamily arXiv:1606.05146}}].
%7 citations counted in INSPIRE as of 10 Dec 2021

\bibitem{Barranco:2017aes}
J.~Barranco, A.~Bernal, J.~C.~Degollado, A.~Diez-Tejedor, M.~Megevand, D.~Nunez, and O.~Sarbach,
\emph{Self-gravitating black hole scalar wigs},
{\emph{Phys. Rev. D} {\bfseries 96} (2017) 024049},
[{{\ttfamily arXiv:1704.03450}}].
%10 citations counted in INSPIRE as of 10 Dec 2021

\bibitem{Huang:2017nho}
Y.~Huang, D.-J.~Liu, X.-H.~Zhai, and X.-Z.~Li,
\emph{Massive charged Dirac fields around Reissner-Nordstr\"om black holes: Quasibound states and long-lived modes},
{\emph{Phys. Rev. D} {\bfseries 96} (2017) 065002},
[{{\ttfamily arXiv:1708.04761}}].
%13 citations counted in INSPIRE as of 10 Dec 2021

\bibitem{Sporea:2019iwk}
C.~A.~Sporea,
\emph{Quasibound states of the Dirac field in Schwarzschild and Reissner\textendash{}Nordstr\"om black hole backgrounds},
{\emph{Mod. Phys. Lett. A} {\bfseries 34} (2019) 1950323},
[{{\ttfamily arXiv:1905.05086}}].
%2 citations counted in INSPIRE as of 10 Dec 2021




\bibitem{kaluza:1921un}
T.~Kaluza, \emph{Zum unit{\"a}tsproblem der physik}, {\emph{Sitzungsber.
  Preuss. Akad. Wiss. Berlin (Math. Phys.)} {\bfseries 27} (1921) 966}.

\bibitem{Klein:1926tv}
O.~Klein, \emph{{Quantum Theory and Five-Dimensional Theory of Relativity. (In
  German and English)}}, {\emph{Z.
  Phys.} {\bfseries 37} (1926) 895}.

\bibitem{ArkaniHamed:1998rs}
N.~Arkani-Hamed, S.~Dimopoulos, and G.~R. Dvali, \emph{{The Hierarchy problem
  and new dimensions at a millimeter}},
 {\emph{Phys. Lett. B}
  {\bfseries 429} (1998) 263},
  [{{\ttfamily arXiv:hep-ph/9803315}}].



\bibitem{Randall:1999ee}
L.~Randall and R.~Sundrum, \emph{{A Large mass hierarchy from a small extra dimension}},
 {\emph{Phys. Rev. Lett.}
  {\bfseries 83} (1999) 3370},
  [{{\ttfamily arXiv:hep-ph/9905221}}].


\bibitem{Antoniadis:1998ig}
I.~Antoniadis, N.~Arkani-Hamed, S.~Dimopoulos, and G.~R.~Dvali,
\emph{{New dimensions at a millimeter to a Fermi and superstrings at a TeV,}}
{\emph{ Phys. Lett. B}
	{\bfseries 436}, (1998) 257},
[{{\ttfamily arXiv:hep-ph/9804398}}].

\bibitem{Randall:1999vf}
L.~Randall and R.~Sundrum, \emph{{An Alternative to compactification}},
 {\emph{Phys. Rev. Lett.}
  {\bfseries 83} (1999) 4690},
  [{{\ttfamily arXiv:hep-th/9906064}}].

\bibitem{Goldberger:1999uk}
W.~D. Goldberger and M.~B. Wise, \emph{{Modulus stabilization with bulk
  fields}}, {\emph{Phys.
  Rev. Lett.} {\bfseries 83} (1999) 4922},
  [{{\ttfamily arXiv:hep-ph/9907447}}].

\bibitem{Gremm:1999pj}
M.~Gremm, \emph{{Four-dimensional gravity on a thick domain wall}},
  {\emph{Phys. Lett. B}
  {\bfseries 478} (2000) 434},
  [{{\ttfamily arXiv:hep-th/9912060}}].

\bibitem{DeWolfe:1999cp}
O.~DeWolfe, D.~Z. Freedman, S.~S. Gubser, and A.~Karch, \emph{{Modeling the
  fifth-dimension with scalars and gravity}},
  {\emph{Phys. Rev. D}
  {\bfseries 62} (2000) 046008},
  [{{\ttfamily arXiv:hep-th/9909134}}].

 \bibitem{Csaki:2000fc}
 C.~Csaki, J.~Erlich, T.~J.~Hollowood, and Y.~Shirman,
 \emph{{Universal aspects of gravity localized on thick branes}},
 {\emph{Nucl. Phys. B} {\bfseries 581}, 309 (2000)},
 [{{\ttfamily arXiv:hep-th/0001033}}].


\bibitem{Bazeia:2008zx}
D.~Bazeia, A.~R. Gomes, L.~Losano, and R.~Menezes, \emph{{Braneworld Models of
  Scalar Fields with Generalized Dynamics}},
  {\emph{Phys. Lett. B}
  {\bfseries 671} (2009) 402},
  [{{\ttfamily arXiv:0808.1815}}].


\bibitem{Charmousis:2001hg}
C.~Charmousis, R.~Emparan, and R.~Gregory, \emph{{Selfgravity of brane worlds: A
  New hierarchy twist}},
  {\emph{JHEP} {\bfseries
  05} (2001) 026}, [{{\ttfamily
  arXiv:hep-th/0101198}}].

\bibitem{Arias:2002ew}
O.~Arias, R.~Cardenas, and I.~Quiros, \emph{{Thick brane worlds arising from
  pure geometry}},
 {\emph{Nucl. Phys. B}
  {\bfseries 643} (2002) 187},
  [{{\ttfamily arXiv:hep-th/0202130}}].

\bibitem{Barcelo:2003wq}
C.~Barcelo, C.~Germani, and C.~F. Sopuerta, \emph{{On the thin shell limit of
  branes in the presence of Gauss-Bonnet interactions}},
  {\emph{Phys. Rev. D}
  {\bfseries 68} (2003) 104007},
  [{{\ttfamily arXiv:gr-qc/0306072}}].

\bibitem{Bazeia:2004dh}
D.~Bazeia and A.~R.~Gomes, \emph{{Bloch brane}}, {\emph{JHEP} {\bfseries 05} (2004) 012}, [{{\ttfamily arXiv:hep-th/0403141}}].

\bibitem{CastilloFelisola:2004eg}
O.~Castillo-Felisola, A.~Melfo, N.~Pantoja, and A.~Ramirez, \emph{{Localizing gravity on exotic thick three-branes}}, {\emph{Phys. Rev. D}
  {\bfseries 70} (2004) 104029},
  [{{\ttfamily arXiv:hep-th/0404083}}].

\bibitem{BarbosaCendejas:2005kn}
N.~Barbosa-Cendejas and A.~Herrera-Aguilar, \emph{{4D gravity localized in non
  $Z_2$ symmetric thick branes}},
  {\emph{JHEP} {\bfseries
  10} (2005) 101}, [{{\ttfamily
  arXiv:hep-th/0511050}}].

\bibitem{Koerber:2008rx}
P.~Koerber, D.~Lust, and D.~Tsimpis, \emph{{Type IIA AdS$_4$ compactifications on
  cosets, interpolations and domain walls}},
  {\emph{JHEP} {\bfseries
  07} (2008) 017}, [{{\ttfamily
  arXiv:0804.0614}}].

\bibitem{BarbosaCendejas:2007vp}
N.~Barbosa-Cendejas, A.~Herrera-Aguilar, M.~A. Reyes~Santos, and C.~Schubert,
  \emph{{Mass gap for gravity localized on Weyl thick branes}},
  {\emph{Phys. Rev. D}
  {\bfseries 77} (2008) 126013},
  [{{\ttfamily arXiv:0709.3552}}].

\bibitem{Johnson:2008kc}
M.~C. Johnson and M.~Larfors, \emph{{Field dynamics and tunneling in a flux
  landscape}}, {\emph{Phys.
  Rev. D} {\bfseries 78} (2008) 083534},
  [{{\ttfamily arXiv:0805.3705}}].

\bibitem{Liu:2011wi}
Y.-X. Liu, Y.~Zhong, Z.-H. Zhao, and H.-T. Li, \emph{{Domain wall brane in
  squared curvature gravity}},
  {\emph{JHEP} {\bfseries 06}
  (2011) 135}, [{{\ttfamily
  arXiv:1104.3188}}].


\bibitem{Kanno:2004nr}
S.~Kanno and J.~Soda, \emph{{Quasi-thick codimension 2 braneworld}},
  {\emph{JCAP} {\bfseries
  0407} (2004) 002}, [{{\ttfamily
  arXiv:hep-th/0404207}}].

\bibitem{Chumbes:2011zt}
A.~E.~R. Chumbes, J.~M. Hoff~da Silva, and M.~B. Hott, \emph{{A model to
  localize gauge and tensor fields on thick branes}},
  {\emph{Phys. Rev. D}
  {\bfseries 85} (2012) 085003},
  [{{\ttfamily arXiv:1108.3821}}].

\bibitem{Andrianov:2012ae}
A.~A. Andrianov, V.~A. Andrianov, and O.~O. Novikov, \emph{{Localization of
  scalar fields on self-gravitating thick branes}},
 {\emph{Phys. Part. Nucl.}
  {\bfseries 44} (2013) 190},
  [{{\ttfamily arXiv:1210.3698}}].

\bibitem{Kulaxizi:2014yxa}
M.~Kulaxizi and R.~Rahman, \emph{{Higher-Spin Modes in a Domain-Wall
  Universe}}, {\emph{JHEP}
  {\bfseries 10} (2014) 193},
  [{{\ttfamily arXiv:1409.1942}}].

\bibitem{Dutra:2014xla}
A.~de~Souza~Dutra, G.~P. de~Brito, and J.~M. Hoff~da Silva, \emph{{Method for
  obtaining thick brane models}},
  {\emph{Phys. Rev. D}
  {\bfseries 91} (2015) 086016},
  [{{\ttfamily arXiv:1412.5543}}].


\bibitem{Chakraborty:2015zxc}
S.~Chakraborty and S.~SenGupta,
\emph{{Kinematics of Radion field: A possible source of dark matter}},
{\emph{Eur. Phys. J. C} {\bfseries 76} (2016) 648},
[{{\ttfamily arXiv:1511.00646}}].

\bibitem{Karam:2018squ}
A.~Karam, A.~Lykkas, and K.~Tamvakis,
\emph{{Frame-invariant approach to higher-dimensional scalar-tensor gravity}},
{\emph{Phys. Rev. D} {\bfseries 97} (2018) 124036},
[{{\ttfamily arXiv:1803.04960}}].







\bibitem{Liu:2009ve}
Y.-X. Liu, J.~Yang, Z.-H. Zhao, C.-E. Fu, and Y.-S. Duan,
\emph{{Fermion Localization and Resonances on A de Sitter Thick Brane}},
 {\emph{Phys. Rev. D} {\bfseries 80} (2009) 065019},
  [{{\ttfamily arXiv:0904.1785}}].


\bibitem{Almeida:2009jc}
C.~A.~S. Almeida, M.~M. Ferreira, A.~R. Gomes, and R.~Casana,
  \emph{{Fermion localization and resonances on two-field thick branes}},
  {\emph{Phys. Rev. D}
  {\bfseries 79} (2009) 125022},
  [{{\ttfamily arXiv:0901.3543}}].

\bibitem{Cruz:2013uwa}
W.~T. Cruz, L.~J.~S. Sousa, R.~V. Maluf, and C.~A.~S. Almeida, \emph{{Graviton
  resonances on two-field thick branes}},
  {\emph{Phys. Lett. B}
  {\bfseries 730} (2014) 314},
  [{{\ttfamily arXiv:1310.4085}}].

\bibitem{Xu:2014jda}
Z.-G. Xu, Y.~Zhong, H.~Yu, and Y.-X. Liu, \emph{{The structure of $f(R)$-brane
  model}}, {\emph{Eur.
  Phys. J. C} {\bfseries 75} (2015) 368},
  [{{\ttfamily arXiv:1405.6277}}].

\bibitem{Csaki:2000pp}
C.~Csaki, J.~Erlich, and T.~J. Hollowood, \emph{{Quasilocalization of gravity by
  resonant modes}},
 {\emph{Phys. Rev. Lett.}
  {\bfseries 84} (2000) 5932},
  [{{\ttfamily arXiv:hep-th/0002161}}].

\bibitem{Zhang:2016ksq}
Y.-P.~Zhang, Y.-Z.~Du, W.-D.~Guo and Y.-X.~Liu,
\emph{{Resonance spectrum of a bulk fermion on branes}},
{\emph{Phys. Rev. D}  {\bfseries93} (2016) 065042},
[{{\ttfamily arXiv:1601.05852}}].


\bibitem{Sui:2020fty}
T.-T. Sui, W.-D. Guo, Q.-Y. Xie and Y.-X. Liu,
\emph{{Generalized geometrical coupling for vector field localization on thick brane in asymptotic Anti-de
  Sitter spacetime}},
 {\emph{Phys. Rev. D}  {\bfseries101} (2020) 055031},
  [{{\ttfamily arXiv:2001.02154}}].



\bibitem{Tan:2020sys}
Q.~Tan, W.-D.~Guo, Y.-P.~Zhang and Y.-X.~Liu,
\emph{Gravitational resonances on $f(T)$-branes},
{\emph{Eur. Phys. J. C} {\bfseries 81} (2021) 373},
[{{\ttfamily arXiv:2008.08440}}].
%3 citations counted in INSPIRE as of 19 Oct 2021

\bibitem{Chen:2020zzs}
J.~Chen, W.-D.~Guo, and Y.-X.~Liu,
\emph{Thick branes with inner structure in mimetic f(R) gravity},
{\emph{Eur. Phys. J. C} {\bfseries 81} (2021) 709},
[{{\ttfamily arXiv:2011.03927}}].
%3 citations counted in INSPIRE as of 10 Dec 2021

\bibitem{Seahra:2005wk}
S.~S.~Seahra,
\emph{Ringing the Randall-Sundrum braneworld: Metastable gravity wave bound states},
{\emph{Phys. Rev. D} {\bfseries  72} (2005) 066002},
[{{\ttfamily arXiv:hep-th/0501175}}].
%20 citations counted in INSPIRE as of 10 Dec 2021

\bibitem{Seahra:2005iq}
S.~S.~Seahra,
\emph{Metastable massive gravitons from an infinite extra dimension},
{\emph{Int. J. Mod. Phys. D} {\bfseries  14} (2005) 2279},
[{{\ttfamily arXiv:hep-th/0505196}}].
%3 citations counted in INSPIRE as of 10 Dec 2021



\bibitem{Yang:2017puy}
K.~Yang, Y.-X.~Liu, B.~Guo, and X.-L.~Du,
\emph{Scalar perturbations of Eddington-inspired Born-Infeld braneworld},
{\emph{Phys. Rev. D}  {\bfseries 96} (2017), 064039},
[{{\ttfamily 1706.04818}}].
%12 citations counted in INSPIRE as of 14 Sep 2022


\bibitem{Wan:2020smy}
J.-J.~Wan, Z.-Q.~Cui, W.-B.~Feng, and Y.-X.~Liu,
\emph{Smooth braneworld in $6$-dimensional asymptotically AdS spacetime},
{\emph{JHEP} {\bfseries  05} (2021), 017},
[{{\ttfamily arXiv:2010.05016}}].
%2 citations counted in INSPIRE as of 14 Sep 2022


\bibitem{Megevand:2007uy}
M.~Megevand, I.~Olabarrieta, and L.~Lehner,
\emph{Scalar field confinement as a model for accreting systems},
{\emph{Class. Quant. Grav.} {\bfseries  24} (2007) 3235},
[{{\ttfamily arXiv:0705.0644}}].
%6 citations counted in INSPIRE as of 13 Dec 2021


\bibitem{Pavlidou:2000cs}
V.~Pavlidou, K.~Tassis, T.~W.~Baumgarte, and S.~L.~Shapiro,
\emph{Radiative falloff in neutron star space-times},
{\emph{Phys. Rev. D} {\bfseries  62} (2000) 084020},
[{{\ttfamily arXiv:gr-qc/0007019}}].
%13 citations counted in INSPIRE as of 13 Dec 2021


\bibitem{Witek:2012tr}
H.~Witek, V.~Cardoso, A.~Ishibashi, and U.~Sperhake,
\emph{Superradiant instabilities in astrophysical systems},
{\emph{Phys. Rev. D} {\bfseries  87} (2013) 043513},
[{{\ttfamily arXiv:1212.0551}}].
%165 citations counted in INSPIRE as of 12 Dec 2021



\bibitem{ATLAS:2017tlw}
M.~Aaboud \textit{et al.} [ATLAS],
\emph{Search for heavy ZZ resonances in the $\ell ^+\ell ^-\ell ^+\ell ^-$ and $\ell ^+\ell ^-\nu \bar{\nu }$ final states using proton\textendash{}proton collisions at $\sqrt{s}= 13$   $\text {TeV}$ with the ATLAS detector},
{\emph{Eur. Phys. J. C} {\bfseries  78} (2018)  293},
[{{\ttfamily arXiv:1712.06386}}].
%137 citations counted in INSPIRE as of 09 Sep 2022

\end{thebibliography}
\end{document}